\documentclass{elsart}
\usepackage{natbib}
\begin{document}
\runauthor{Lattimer, James M}
\begin{frontmatter}
\title{Nuclear Matter and its Role in Supernovae, Neutron Stars
and Compact Object Binary Mergers\thanksref{X}}
\author[SUNYSB]{James M. Lattimer} and
\author[SUNYSB]{Madappa Prakash}
\thanks[X]{Partially supported by USDOE Grants DE-AC02-87ER40317 and
DE-FG02-88ER-40388,
and by NASA ATP Grant \# NAG 52863.}
\address[SUNYSB]{Dept. of Physics \& Astronomy, State University of New York at
Stony Brook,
Stony Brook, NY 11794-3800}
\begin{abstract}
The equation of state (EOS) of dense matter plays an important role in the
supernova phenomenon, the structure of neutron stars, and in the
mergers of compact objects (neutron stars and black holes).  During
the collapse phase of a supernova, the EOS at subnuclear
densities controls the collapse rate, the amount of deleptonization
and thus the size of the collapsing core and the bounce density.
Properties  of nuclear matter that are especially crucial are the
symmetry energy and the nuclear specific heat.  The nuclear
incompressibility, and the supernuclear EOS, play
supporting roles.  In a similar way, although the maximum masses of
neutron stars are entirely dependent upon the supernuclear EOS,
other important structural aspects are more sensitive to the
equation of state at nuclear densities.  The radii, moments of
inertia, and the relative binding energies of neutron stars are, in
particular, sensitive to the behavior of the nuclear symmetry energy.
The dependence of the radius of a neutron star on its mass is shown to
critically influence the outcome of the compact merger of two neutron
stars or a neutron star with a small mass black hole.  This latter
topic is especially relevant to this volume, since it stems from
research prompted by the tutoring of David Schramm a quarter century
ago.\end{abstract}
\end{frontmatter}
\section{Introduction}
The equation of state (EOS) of dense matter plays an important role in
the supernova phenomenon and in the structure and evolution of neutron
stars.  Matter in the collapsing core of a massive star at the end of
its life is compressed from white dwarf-like densities of about $10^6$
g cm$^{-3}$ to two or three times the nuclear saturation density,
about $3\cdot10^{14}$ g cm$^{-3}$ or $n_s=0.16$ baryons fm$^{-3}$. The
central densities of neutron stars may range up to 5--10 $n_s$.  At
densities around $n_s$ and below matter may be regarded as a mixture
of neutrons, protons, electrons and positrons, neutrinos and
antineutrinos, and photons.  At higher densities, additional
constituents, such as hyperons, kaons, pions and quarks may be
present, and there is no general consensus regarding the properties of
such ultradense matter.  Fortunately for astrophysics, however, the
supernova phenomenon and many aspects of neutron star structure may
not depend upon ultradense matter, and this article will focus on the
properties of matter at lower densities.

The main problem is to establish the state of the nucleons, which may
be either bound in nuclei or be essentially free in continuum states.
Neither temperatures nor densities are large enough to excite
degrees of freedom, such as hyperons, mesons or quarks.
Electrons are rather weakly interacting and may be treated as an ideal
Fermi gas: at densities above $10^7$ g cm$^{-3}$, they are
relativisitic.  Because of their even weaker interactions, photons and
neutrinos (when they are confined in matter) may also be treated as
ideal gases.

At low enough densities and temperatures, and provided the matter does
not have too large a neutron excess, the relevant nuclei are stable in
the laboratory, and experimental information may be used directly.
The so-called Saha equation may be used to determine their relative
abundances.  Under more extreme conditions, there are a number of
important physical effects which must be taken into account.  At
higher densities, or at moderate temperatures, the neutron chemical
potential increases to the extent that the density of nucleons outside
nuclei can become large.  It is then important to treat matter outside
nuclei in a consistent fashion with that inside.  These nucleons will
modify the nuclear surface, decreasing the surface tension.  At finite
temperatures, nuclear excited states become populated, and these
states can be included by treating nuclei as warm drops of nuclear
matter.  At low temperatures, nucleons in nuclei are degenerate and
Fermi-liquid theory is probably adequate for their description.
However, near the critical temperature above which the dense phase of
matter inside nuclei can no longer coexist with the lighter phase of
matter outside nuclei, the equilibrium of the two phases of matter is
crucial.

The fact that at subnuclear densities the spacing between nuclei may
be of the same order of magnitude as the nuclear size itself will lead
to substantial reductions in the nuclear Coulomb energy.  Although
finite temperature ``plasma'' effects will modify this, the
zero-temperature Wigner-Seitz approximation employed by Baym, Bethe
\& Pethick~\cite{BBP} is usually adequate.  Near the nuclear
saturation density, nuclear deformations must be dealt with, including
the possibilities of ``pasta-like'' phases and matter turning
``inside-out'' ({\it i.e.}, the dense nuclear matter envelopes a
lighter, more neutron-rich, liquid).  Finally, the translational
energy of the nuclei may be important under some conditions.  This
energy is important in that it may substantially reduce the average
size of the nuclear clusters.

An acceptable way of bridging the regions of low density and
temperature, in which the nuclei can be described in terms of a simple
mass formula, and high densities and/or high temperatures in which the
matter is a uniform bulk fluid, is to use a compressible liquid
droplet model for nuclei in which the drop maintains thermal,
mechanical, and chemical equilibrium with its surroundings.  This
allows us to address both the phase equilibrium of nuclear matter,
which ultimately determines the densities and temperatures in which
nuclei are permitted, and the effects of an external nucleon fluid on
the properties of nuclei.  Such a model was originally developed by
Lattimer {\it et al.\/} \cite{LPRL} and modified by Lattimer \&
Swesty~\cite{LS}.  This work was a direct result of David Schramm's
legendary ability to mesh research activities of various groups, in
this case to pursue the problem of neutron star decompression.  After
the fact, the importance of this topic for supernovae became apparent.

\section{Nucleon Matter Properties}

The compressible liquid droplet model rests upon
the important fact that in a many-body system
the nucleon-nucleon interaction exhibits saturation.
Empirically, the energy per particle of bulk nuclear matter
reaches a minimum, about --16 MeV, at a density
$n_s\cong 0.16~{\rm fm}^{-3}$.
Thus, close to $n_s$, its
density dependence is approximately parabolic.  The nucleon-nucleon
interaction is optimized for equal numbers of neutrons and protons
(symmetric matter), so a parabolic dependence on the neutron
excess or proton fraction, $x$, can be assumed.  About a third to a
half of the energy
change made by going to asymmetric matter is due to the nucleon
kinetic energies, and to a good approximation, this varies as
$(1-2x)^2$ all the way to pure neutron matter ($x=0$).  The $x$
dependence of the potential terms in most theoretical models can also
be well approximated by a quadratic dependence.
Finally, since at low temperatures the
nucleons remain degenerate, their temperature dependence to
leading order is also quadratic.  Therefore, for analytical purposes,
the nucleon free energy per baryon can be approximated as
$f_{bulk}(n,x)$, in MeV, as
\begin{eqnarray}
f_{bulk}(n,x)\simeq-16+S_v(n)(1-2x)^2+{K_s\over18}\Biggl({n\over
n_s}-1\Biggr)^2\cr
-{K^\prime_s\over27}\Biggl({n\over
n_s}-1\Biggr)^3-a_v(n,x)T^2 \,,\label{bulk}
\end{eqnarray}
where $a_v(n)=(2m^*/\hbar^2)(\pi/12n)^{2/3}$.  The expansion
parameters, whose values are uncertain to varying degrees, are the
incompressibility, $K_s=190-250$ MeV, the skewness parameter
$K_s^\prime=1780-2380$ MeV, the symmetry energy coefficient $S_v\equiv
S_v(n_s)=25-36$ MeV, and the bulk level density parameter,
$a_v(n_s,x=1/2)\simeq(1/15)(m^*(n_s,x=1/2)/m)$ MeV$^{-1}$, where $m^*$
is the effective mass of the nucleon.  Values for $m^*(n_s,x=1/2)/m$
are in the range $0.7-0.9$.  The general definition of the
incompressibility is $K=9dP/dn=9d(n^2df_{bulk}/dn)/dn$, where $P$ is
the pressure, and $K_s\equiv K(n_s,1/2)$.  It is worthwhile noting
that the symmetry energy and nucleon effective mass (which directly
affects the matter's specific heat) are density dependent, but these
dependencies are difficult to determine from experiments. The
parameters, and their density dependences, characterize the nuclear
force model and are essential to our understanding of astrophysical
phenomena.

The experimental determination of these parameters has come from
comparison of the total masses and energies of giant resonances of laboratory
nuclei with theoretical predictions.  Some of these comparisons are
easily illustrated with the compressible liquid droplet model.  In
this model, the nucleus is treated as uniform drop of nuclear matter
with temperature $T$, density $n_i$ and proton fraction $x_i$.  The
nucleus will, in general, be surrounded by and be in equilibrium with
a vapor of matter with density $n_o$ and proton fraction $x_o$.  At
low ambient densities $n$ and vanishing temperature, the outside vapor
vanishes.  Even at zero temperature, if $n$ is large enough, greater
than the so-called neutron drip density $n_d\simeq1.6\cdot10^{-3}$
fm$^{-3}$, the neutron chemical potential of the nucleus is positive
and ``free'' neutrons exist outside the nucleus.  At finite
temperature, the external vapor consists of both neutrons and protons.
In addition, because of their high binding energy, $\alpha-$particles
will also be present.  The total free energy density is the sum of the
various components:
\begin{equation}
F=F_H+F_o+F_\alpha+F_e+F_\gamma\,.
\end{equation}
Here, $F_H$ and $F_o$ represent the free energy densities of the heavy
nuclei and the outside vapor, respectively.  The energy densities of
the electrons and photons, $F_e$ and $F_\gamma$, are independent of
the baryons and play no role in the equilibrium. For simplicity, we
neglect the role of $\alpha$-particles in the following discussion
(although it is straightforward to include their effect~\cite{LPRL}).

In the compressible liquid drop model, it is assumed that the nuclear
energy can be written as an expansion in $A^{1/3}$ and $(1-2x_i)^2$:
\begin{equation}
F_H=un_i[f_{bulk}+f_{surf}+f_{Coul}+f_{trans}]\,,\label{ftot}
\end{equation}
where the $f$'s represent free energies per baryon due to the bulk,
surface, Coulomb, and translation, respectively.  The bulk energy, for
example, is given by Eq.~(\ref{bulk}).  The surface energy can be
parametrized as
\begin{equation}f_{surf}=4\pi R^2\sigma(x_i,T)\equiv4\pi R^2h(T)[\sigma_o-\sigma_s(1-2x_i)^2]\,,\label{fsurf}
\end{equation}
where $R$ is the nuclear charge radius,
$h(T)$ is a calculable function of temperature,
$\sigma_o$ is the surface tension of symmetric matter, 
and $\sigma_s=(n_i^2/36\pi)^{1/3}S_s$ where $S_s$ is the
surface symmetry energy coefficient from the traditional mass formula.  In
this simplified discussion, the influence of the neutron
skin~\cite{LPRL}, which distinguishes the ``drop model'' from the ``droplet
model'', is omitted.  The Coulomb energy, in the Wigner-Seitz
approximation~\cite{BBP}, is
\begin{equation}
f_{Coul}=0.6x_i^2A^2e^2D(u)/R\,,\label{fcoul}
\end{equation}
where $D(u)=1-1.5u^{1/3}+0.5u$ and $u$ is the fraction of the
volume occupied by nuclei.  If the fractional mass of matter outside
the nuclei is small, $u\simeq n/n_i$.

It is clear that additional parameters, $S_s$ and another involving
the temperature dependence of $h$, exist in conjunction with those
defining the expansions of the bulk energy.  The temperature
dependence is related to the matter's critical temperature $T_c$ at
which the surface disappears.  It is straightforward to demonstrate
from the thermodynamic relations defining $T_c$, namely $\partial
P_{bulk}/\partial n=0$ and $\partial^2 P_{bulk}/\partial n^2=0$, that
$T_c\propto \sqrt{K_s}$.  Therefore, the specific heat to be
associated with the surface energy will in general be proportional to
$T_c^{-2}\propto K_s^{-1}$.  About half the total specific
heat originates in the surface, so $K_s$ influences the temperature
for a given matter entropy, important during stellar collapse.

The equilibrium between nuclei and their surroundings is determined by
minimizing $F$ with respect to its internal variables, at fixed
$n,Y_e$, and $T$.  This is described in more detail in Refs.~\cite{LPRL,
LS}, and leads to equilibrium conditions involving the pressure and the
baryon chemical potentials, as well
as a condition determining the nuclear size $R$.  The latter is analogous
to the one found by Baym, Bethe \& Pethick~\cite{BBP} who equated
the nuclear surface energy with twice the Coulomb energy.  The
relations in Eqs.~(\ref{fsurf}) and (\ref{fcoul}) lead to
\begin{equation}
R=\Biggl[{15\sigma(x_i)\over8\pi e^2 x_i^2
n_i^2}\Biggr]^{1/3}\,.
\end{equation}
Experimental limits to $K_s$, most importantly from RPA analyses of
the breathing mode of the giant monopole resonance~\cite{Blaizot}, give
$K_s\cong230$ MeV.  It is also possible to obtain values from the
so-called scaling model developed from the compressibile liquid drop
model.  The finite-nucleus incompressibility is
\begin{equation}
K(A,Z)=(M/\hbar^2)R^2E^2_{br}\,,\end{equation} where $M$ is the
mass of the nucleus and $E_{br}$ is the breathing-mode
energy.  $K(A,Z)$ is commonly expanded  as
\begin{equation}
K(A,Z)=K_s+K_{surf}A^{-1/3}+K_{vI}I^2+K_{surfI}I^2A^{-1/3}
+K_CZ^2A^{-4/3}\,,\label{compaz}\end{equation} and then fit by
least squares to the data for $E_{br}$.  Here the asymmetry
$I=1-2Z/A$.  For a given assumed value of $K_s$, and taking
$K_{surfI}=0$, Pearson~\cite{P}
showed that experimental data gave
\begin{eqnarray}K_C\simeq15.4-0.065 K_s\pm2 {\rm~MeV}\,,\quad
K_{surf}\simeq230-3.2 K_s\pm50 {\rm~MeV}\,.\label{constr}\end{eqnarray}
With minimal assumptions regarding the form of the nuclear force,
Pearson~\cite{P} demonstrated that values of $K_s$ ranging from
200 MeV to more than 350 MeV could be consistent with experimental
data.

But the liquid drop model predicts other relations between the parameters:
\begin{eqnarray}K(A,Z)&=&R^2{\partial^2 E(Z,A)/A\over\partial
R^2}\Biggr|_A=9n^2{\partial^2E(Z,A)/A\over\partial n^2}\Biggr|_A\,,\cr
0=P(A,Z)&=&R{\partial^2 E(Z,A)/A\over\partial R}\Biggr|_A=3n{\partial
E(Z,A)/A\over\partial n}\Biggr|_A\,.
\end{eqnarray}
Here $E(Z,A)$ is the total energy of the nucleus, and is equivalent to
Eq.~(\ref{ftot}).  The second of these equations simply expresses the
equilibrium between the nucleus and the surrounding vacuum, which
implies that the pressure of the bulk matter inside the nucleus is
balanced by the pressure due to the curvature of the surface and the
Coulomb energy.  It can then be shown that
\begin{eqnarray}
K_C&=&-(3e^2/5r_o)[8+27n_s^3f_{bulk}^{\prime\prime\prime}(n_s)/K_s]\,,\cr
K_{surf}&=&4\pi r_o^2\sigma_o[9n_s^2\sigma_o^{\prime\prime}/\sigma_o+22+
54n_s^3f_{bulk}^{\prime\prime\prime}(n_s)/K_s]\,,\cr
K_{surfI}&=&4\pi r_o^2\sigma_s[9n_s^2\sigma_s^{\prime\prime}/\sigma_s+22+
54n_s^3f_{bulk}^{\prime\prime\prime}(n_s)/K_s]\,,\cr
K_I&=&9[n_s^2S_v^{\prime\prime}(n_s)-2n_oS_v^\prime(n_s)-9n_s^4S_v^\prime(n_s)
f_{bulk}^{\prime\prime\prime}(n_s)/K_s]\,.\label{dropcomp}
\end{eqnarray}
Primes denote derivatives with respect to the density.  From these
relations, and again assuming $K_{surfI}=0$, Pearson demonstrated that
an interesting correlation between $K_s$ and $K^\prime_s$, where
$K^\prime_s\equiv-27n_s^3f_{bulk}^{\prime\prime\prime}(n_s)$, could be
obtained:
\begin{equation}
K^\prime_s=-0.0860K_s^2+(28.37\pm2.65)K_s\,.\label{comp}\end{equation}
Assuming $K_s\simeq190-250$ MeV, this suggests that
$K_s^\prime=1780-2380$ MeV, a potential constraint.
Alternatively, eliminating $K_s^\prime$, one finds
\begin{equation}
K_s=137.4-26.36n_s^2\sigma_o^{\prime\prime}/\sigma_o\pm23.2
{\rm~MeV}\,.\end{equation} The second derivative of the surface
tension can be deduced from Hartree-Fock or Thomas-Fermi semi-infinite
surface calculations.  For example, if a parabolic form of $f_{bulk}$
is used, one finds
\begin{equation}
n_s^2\sigma_o^{\prime\prime}/\sigma_o=-6\end{equation} leading to
$K_s=295.5\pm23.2$ MeV.  In general, the density dependence of $S_v$
will decrease the magnitudes of $K_s$ and $\sigma_o^{\prime\prime}$
from the above values.

It is hoped current experimental work will tighten these constraints.
A shortcoming of the scaling model is that, to date, the surface
symmetry energy term was neglected.  This is not required, however,
and further work is necessary to resolve this matter.

Because the surface energy  represents the energy difference between uniformly
and realistically distributed nuclear material in a nucleus, the
parameter $S_s$ can be related to the density dependence of $S_v(n)$ and
to $K_s$.  If $f_{bulk}$ is assumed to behave quadratically with
density around $n_s$, this relation can be particularly simply
expressed~\cite{L}:
\begin{equation}
{S_s\over S_v}={3\over\sqrt{2}}{a_{1/2}\over r_o}\int_0^1
{\sqrt{x}\over1-x}\Biggl[{S_v\over S_v(xn_s)}-1\Biggr]dx. \end{equation}
Here, $S_v\equiv S_v(n_s)$, $a_{1/2}=(dr/d\ln n)_{n_s/2}$ is a measure
of the thickness of the nuclear surface and $r_o=(4\pi
n_s/3)^{-1/3}=R/A^{1/3}$.  If $S_v(n)$ is linear, then the integral is
2; if $S_v(n)\propto n^{2/3}$, then the integral is 0.927.  Since
$a_{1/2}$ will be sensitive to the value of $K_s$, we expect the value
of $S_s/S_v$ to be also.

Experimentally, there are two major sources of information regarding
the symmetry energy parameters: nuclear masses and giant resonance energies.
However, because of the small excursions in $A^{1/3}$ afforded  by
laboratory nuclei, each source provides only a correlation between
$S_s$ and $S_v$.  For example, the total symmetry energy in
the liquid droplet model (now explicitly including the presence of the
neutron skin, see Ref.~\cite{LPRL}) is
\begin{equation}
E_{sym}=(1-2x_i)^2S_v/[1+(S_s/S_v)A^{-1/3}].
\end{equation}
Evaluating $\alpha=d \ln S_s/d \ln S_v$ near the ``best-fit''
values $S_{s0}$ and $S_{v0}$, one finds
\begin{equation}
\alpha\simeq2+S_{v0}<A>^{1/3}/S_{s0}\simeq6\,,\label{ssfit}
\end{equation}
where $<A>^{1/3}$ for the fitted nuclei is about 5.  Thus, as the
value of $S_v$ is changed in the mass formula, the value of $S_s$ must
vary rapidly to compensate.

An additional correlation between these parameters can be obtained
from the fitting of isovector giant resonances, and this has the
potential of breaking the degeneracy of $S_v$ and $S_s$, because it
has a different slope~\cite{L}.  Lipparini \& Stringari~\cite{Lip}
used a hydrodynamical model of the nucleus to derive the isovector
resonance energy:
\begin{eqnarray}
E_d&=&\sqrt{{24\hbar^2\over m^*}{NZ\over A} \Bigl[\int
{nr^2S_v\over S_v(n)} d^3r\Bigr]^{-1}}\cr&\simeq&96.5\sqrt{{m\over
m^*}
{S_v\over30{\rm~MeV}}\Biggl[1+{5S_s\over3S_vA^{1/3}}\Biggr]^{-1}}A^{-1/3}
{\rm~MeV},
\label{lsdip}
\end{eqnarray}
where $m^*$ is an effective nucleon mass.  This relation results in a
slightly less-steep correlation between $S_s$ and $S_v$,
\begin{equation}
\alpha=2/m^*+(3/5)S_{v0}<A>^{1/3}/S_{s0}\simeq4-5\,.
\end{equation}
Unfortunately the value
of $m^*$ is an undetermined parameter and this slope is not very
different from that obtained from fitting masses.  Therefore,
uncertainties in the model make a large difference to the crossing
point of these two correlations.  A strong theoretical attack,
perhaps using further RPA analysis,
together with more experiments to supplement
the relatively meager amount of existing data, would be very useful.

\section{The Equation of State and the Collapse of Massive Stars}

Massive stars at the end of their lives are believed to consist of a
white dwarf-like iron core of 1.2--1.6 M$_\odot$ having low entropy
($s\le1$), surrounded by layers of less processed material from shell
nuclear burning.  The effective Chandrasekhar mass, the maximum mass
the degenerate electron gas can support, is dictated by the entropy
and the average lepton content, $Y_L$, believed to be around
0.41--0.43.  As mass is added to the core by shell Si-burning, the
core eventually becomes unstable and collapses.

During the collapse, the lepton content decreases due to net electron
capture on nuclei and free protons.  But when the core density
approaches $10^{12}$ g cm$^{-3}$, the neutrinos can no longer escape
from the core on the dynamical collapse time~\cite{Sato}.  After
neutrinos become trapped, $Y_L$ is frozen at a value of about
0.38--0.40, and the entropy is also thereafter fixed.  The core
continues to collapse until the rapidly increasing pressure reverses
the collapse at a bounce density of a few times nuclear density.

The immediate outcome of the shock generated by the bounce is also
dependent upon $Y_L$.  First, the shock energy is determined by the
net binding energy of the post-bounce core, and is proportional to
$Y_e^{10/3}$~\cite{LBY}.  Second, the shock is largely dissipated by
the energy required to dissociate massive nuclei in the
still-infalling matter of the original iron core outside the
post-bounce core.  The larger the $Y_L$ of the core, the larger its
mass and the smaller this shell.  Therefore, the progress of the shock
is very sensitive to the value of $Y_L$.

The final value of $Y_L$ is controlled by weak interaction rates, and
is strongly dependent upon the fraction of free protons, $X_p$, which
is proportional to $\exp (\mu_p/T)$, and the phase space available for
proton capture on nuclei, which is proportional to $\mu_e-\hat\mu$, where
$\hat\mu=\mu_n-\mu_p$.  Both are sensitive to the proton fraction in
nuclei ($x_i$) and are largely controlled by $Y_L$.  In addition, the
specific heat controls the temperature which has a direct
influence upon the free proton abundance and the net electron capture
rate.  In spite of the intricate feedback, nuclear parameters
relating chemical potentials to composition, especially $S_v$ and
$S_s$, are obviously important.

\begin{figure}[h]
\vspace{24pc}
\includegraphics{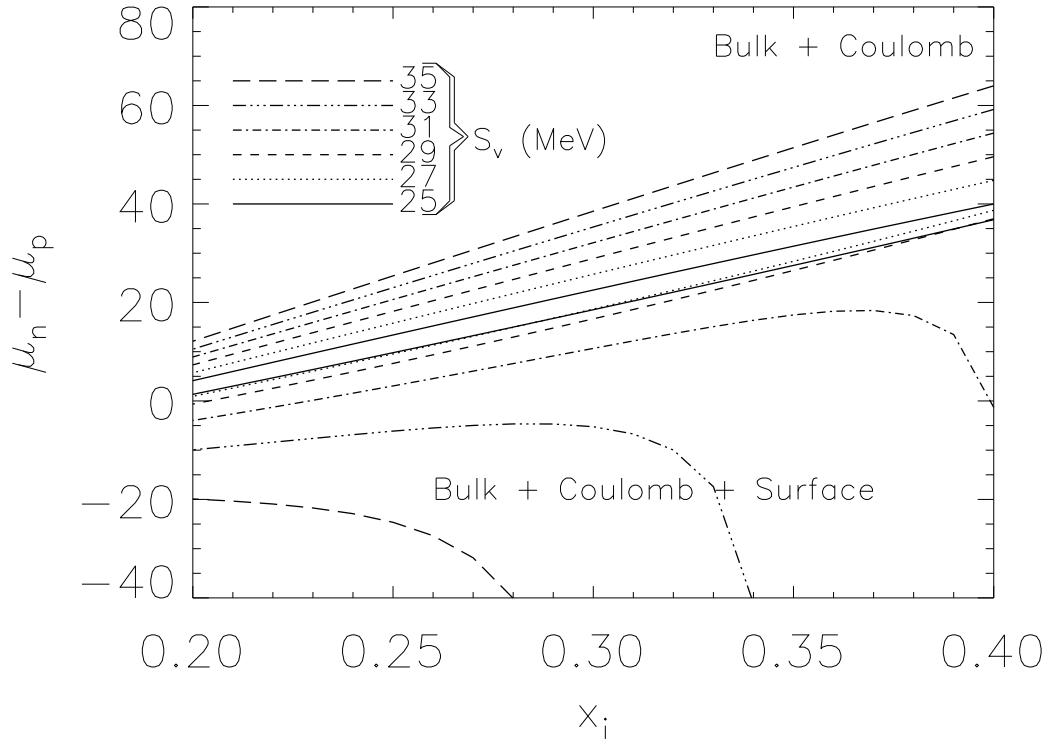}
\caption{Comparison of $\hat\mu=\mu_n-\mu_p$ as a function of $x_i$
for various assumed values of $S_v$, both including and excluding the
effects of the surface symmetry energy.}
\label{muhat}
\end{figure}

As an example, consider
$\hat\mu=\mu_n-\mu_p=-{n_i}^{-1}\partial F_H/\partial x_i$.
With the model of Eqs.~(\ref{ftot})-(\ref{fcoul}), one has
\begin{equation}\hat\mu=4S_v(1-2x_i)-\Biggl({72\pi e^2D\over5x_in_i}\Biggr)^{1/3}
{\sigma_o-\sigma_s(1-2x_i)(1-6x_i)\over(\sigma_o-\sigma_s(1-2x_i)^2)^{1/3}}\,.
\label{muhate}\end{equation}
Recall that $\sigma_s\propto S_s$.  Although the bulk and Coulomb terms alone
(Eq.~\ref{muhate} with $\sigma_s=0$) imply that $\hat\mu$ for a given
$x_i$ rises with increasing $S_v$, the proper inclusion of the surface
symmetry energy gives rise to the opposite behavior.  This is
illustrated in Fig.~\ref{muhat}.

\begin{figure}[h]
\vspace{25pc}
\includegraphics{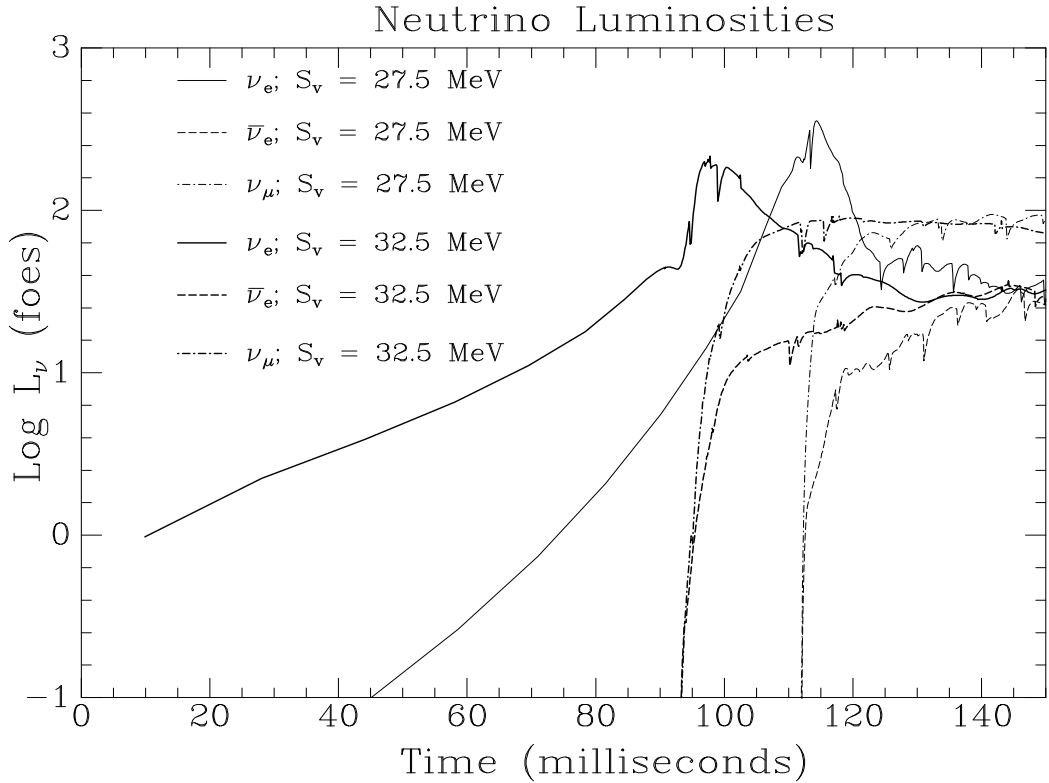}
\caption{The neutrino luminosities during infall as a function of the
bulk symmetry energy parameter.}
\label{nulum}
\end{figure}

Uncertainties in nuclear parameters can thus be expected to have an
influence upon the collapse of massive stars, for example, in the
collapse rate, the final trapped lepton fraction, and the radius at
which the bounce-generated shock initially stalls.  Swesty, Lattimer
\& Myra~\cite{SLM} investigated the effects upon stellar collapse of
altering parameters in a fashion constrained by nuclear systematics.
They found that as long as the parameters permitted a neutron star
maximum mass above the PSR1913+16 mass limit (1.44 M$_\odot$), the
shock generated by core bounce consistently stalls near 100 km,
independently of the assumed $K_s$ in the range 180--375 MeV and $S_v$
in the range 27--35 MeV.  Ref.~\cite{SLM} also found that the final
trapped lepton fraction is also apparently independent of variations
in both $K_s$ and $S_v$.  These results are in contrast to earlier
simulations which had used EOSs that could not support
cold, catalyzed 1.4 M$_\odot$ stars, or in which $S_s$ was not varied
consistently with $S_v$. The strong feedback between the EOS,
weak interactions, neutrino transport, and hydrodynamics is
an example of {\em Mazurek's Law}.

In fact, the only significant consequence of varying $S_v$ involved the
pre-bounce neutrino luminosities.  Increasing
$S_v$ increases the electron capture rate (proportional to
$\mu_e-\hat\mu$ and therefore increases the $\nu_e$ luminosity during
collapse, as shown in Fig.~\ref{nulum}.  Nevertheless, the collapse
rate also increases, so that neutrino trapping occurs sooner and the
final trapped lepton fraction does not change.  It is possible that
large neutrino detectors such as Super-Kamiokande or SNO may be able
to observe an enhanced early rise in neutrino luminosity from nearby
galactic supernovae.

\section{The Structure of Neutron Stars}

The theoretical study of the structure of neutron stars is crucial if
new observations of masses and radii are to lead to effective
constraints on the EOS of dense matter.  This study becomes ever more
important as laboratory studies may be on the verge of yielding
evidence about the composition and stiffness of matter beyond $n_s$.
To date, several accurate mass determinations of neutron stars are
available, and they all lie in a narrow range ($1.25-1.44$ M$_\odot$).
There is some speculation that the absence of neutron stars with
masses above 1.5 M$_\odot$ implies that $M_{max}$ for neutron stars
has approximately this value.  However, since fewer than 10 neutron
stars have been weighed, and all these are in binaries, this
conjecture is premature. Theoretical studies of dense matter
indicate that considerable uncertainties exist in
the high-density behavior of the EOS largely because of the poorly
constrained many-body interactions.
These uncertainties are reflected in a significant
uncertainty in the maximum mass of a beta-stable neutron star, which
ranges from 1.5--2.5 M$_\odot$.

\begin{figure}[h]
\vspace{23pc}
\includegraphics{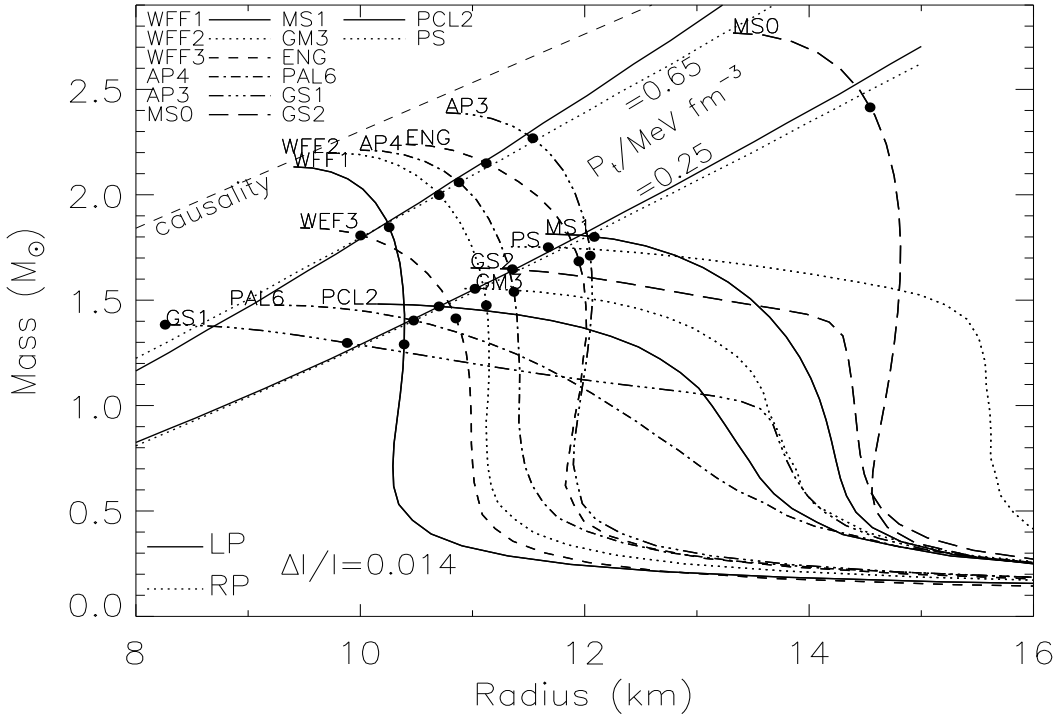}
\caption{$M-R$ curves for the EOSs listed in Table 1.   The diagonal
lines represent two theoretical estimates (LP=Ref.~\cite{LP};
RP=Ref.~\cite{RP}) of the locus of points for $\Delta I/I=1.4\%$ for
extremal limits of $P_t$, 0.25 and 0.65 MeV fm$^{-3}$.  The large dots
on the $M-R$ curves are the exact results.  The region to the left of
the contours labeled 0.65 is not allowed if current glitch models are
correct~\cite{Link}.}
\label{fig:M-R}
\end{figure}

There is some theoretical support for a
lower mass limit for neutron stars in the range $1.1-1.2$ M$_\odot$.
This follows from the facts that the collapsing core of a massive star
is always greater than 1 M$_\odot$ and the minimum mass of a
protoneutron star with a low-entropy inner core of $\sim0.6$ M$_\odot$
and a high-entropy envelope is at least 1.1 M$_\odot$.
Observations from the Earth of thermal radiation from neutron star
surfaces could yield values of the quantity
$R_\infty=R/\sqrt{1-2GM/Rc^2}$, which results from redshifting the
stars luminosity and temperature.  
$M-R$ trajectories for representative EOSs (discussed below) 
are shown in
Figure \ref{fig:M-R}.  
It appears difficult to simultaneously have $M>1$M$_\odot$ and $R_\infty < 
12$ km. 
Those pulsars with at least some suspected thermal
radiation generically yield effective values of $R_\infty$ so small
that it is believed that the radiation originates from polar hot spots
rather than from the surface as a whole.  Other attempts to deduce a
radius include analyses~\cite{Tit} of X-ray bursts from sources 4U
1705-44 and 4U 1820-30 which implied rather small values,
$9.5<R_\infty<14$ km.  However, the modeling of the photospheric
expansion and touchdown on the neutron star surface requires a model
dependent relationship between the color and effective temperatures,
rendering these estimates uncertain.  Absorption lines in X-ray
spectra have also been investigated with a view to deducing the
neutron star radius.  Candidates for the matter producing the
absorption lines are either the accreted matter from the companion
star or the products of nuclear burning in the bursts.  In the former
case, the most plausible element is thought to be Fe, in which case
the relation $R\approx3.2GM/c^2$, only slightly larger than the
minimum possible value based upon causality,~\cite{LPMY,glen} is
inferred.  In the latter case, plausible candidates are Ti and Cr, and
larger values of the radius would be obtained.  In both cases, serious
difficulties remain in interpreting the large line widths, of order
100--500 eV, in the $4.1 \pm 0.1$ keV line observed from many sources.
A first attempt at using light curves and pulse fractions from pulsars
to explore the $M-R$ relation suggested relatively large radii, of
order 15 km~\cite{Page}.  However, this method, which assumed dipolar
magnetic fields, was unable to satisfactorily reconcile the calculated
magnitudes of the pulse fractions and the shapes of the light curves
with observations.

Prospects for a radius determination have improved in recent years,
however, with the detection of a nearby neutron star, RX J185635-3754,
in X-rays and optical radiation~\cite{Walter}. The observed X-rays,
from the ROSAT satellite, are consistent with blackbody emission with
an effective temperature of about 57 eV and very little extinction.
In addition, the fortuitous location of the star in the foreground of
the R CrA molecular cloud limits the distance to $D<120$ pc.  The fact
that the source is not observable in radio and its lack of variability
in X-rays implies that it is not a pulsar unlike other identified
radio-silent isolated neutron stars.  This gives the hope that the
observed radiation is not contaminated with non-thermal emission as is
the case for pulsars.  The X-ray observations of RXJ185635-3754 alone
yield $R_\infty\approx 7.3 (D/120 {\rm~pc}){\rm~ km}$ for a best-fit
blackbody.  Such a value is too small to be consistent with any
neutron star with more than 1 M$_\odot$.  But the optical flux is
about a factor of 2.5 brighter than what is predicted for the X-ray
blackbody, which is consistent with there being a heavy-element
atmosphere~\cite{Romani}.  With such an atmosphere, it is
found~\cite{ALPW} that the effective temperature is reduced to
approximately 50 eV and $R_\infty$ is also increased, to a value of
approximately $21.6 (D/120 {\rm~pc}){\rm~ km}$.  Upcoming parallax
measurements with the Hubble Space Telescope should permit a distance
determination to about 10-15\% accuracy.  If X-ray spectral features
are discovered with the planned Chandra and XMM space observatories,
the composition of the neutron star atmosphere can be inferred, and
the observed redshifts will yield independent mass and radius
information.  In this case, {\em both} the mass and radius of this
star will be found.

Furthermore, a proper motion of 0.34 $^{\prime\prime}$ yr$^{-1}$ has
been detected, in a direction that is carrying the star away from the
Upper Scorpius (USco) association~\cite{ALPW}.  With an assumed
distance of about 80 pc, the positions of RX J185635-3754 and this
association overlap about 800,000 years ago.  The runaway
OB star $\zeta$ Oph is also moving away from USco, appearing to have
been ejected on the order of a million years ago.  The superposition
of these three objects is interesting, and one can speculate that this is
not coincidental.  If upcoming parallax measurements are consistent
with a distance to RX J185635-3754 of about 80 pc, the evidence for
this scenario will be strong, and a good age estimate will result.

In this section, a striking empirical relationship is noted which
connects the radii of neutron stars and the pressure of matter in the
vicinity of $n_s$.  In addition, a number of analytic, exact,
solutions to the general relativistic TOV equation of hydrostatic
equilibrium are explored that lead to several useful approximations
for neutron star structure which directly correlate observables such as
masses, radii, binding energies, and moments of inertia.  The binding
energy, of which more than 99\% is carried off in neutrinos, will be
revealed from future neutrino observations of supernovae.  Moments of
inertia are connected with glitches observed in the spin down of
pulsars, and their observations yield some interesting conclusions
about the distribution of the moment of inertia within the rotating
neutron star.  From such comparisons, it may become easier to draw
conclusions about the dense matter EOS when firm observations of
neutron star radii or moments of inertia become available to accompany
the several known accurate mass determinations.

\subsection{Neutron Star Radii}

\begin{table*}
\caption{Equations of state used in this work.  Approach refers to the
basic theoretical paradigm.  Composition refers to strongly
interacting components (n=neutron, p=proton, H=hyperon, K=kaon,
Q=quark); all approaches include leptonic contributions.}
\vspace*{0.1in}
\begin{center}
\begin{tabular}{l|l|l|l} \hline\hline
Symbol & Reference & Approach & Composition \\ \hline
FP  & \cite{FP} & Variational & np \\
PS & \cite{PS} & Potential & n$\pi^0$ \\
WFF(1-3) & \cite{WFF} & Variational & np \\
AP(1-4) & \cite{Akmal} & Variational & np \\
MS(1-3) & \cite{MS} & Field Theoretical & np \\   
MPA(1-2) & \cite{MPA} & Dirac-Brueckner HF & np \\
ENG & \cite{Engvik} & Dirac-Brueckner HF & np \\
PAL(1-6)  & \cite{PAL} & Schematic Potential & np \\
GM(1-3) & \cite{GM} & Field Theoretical & npH \\
GS(1-2) & \cite{GS} & Field Theoretical & npK\\
PCL(1-2) & \cite{PCL} & Field Theoretical & npHQ
\\ 
SQM(1-3) & \cite{PCL} & Quark Matter & Q $(u,d,s)$\\
\hline
\end{tabular}
\label{eosname}
\end{center}
\end{table*}

The composition of a neutron star chiefly depends on the nature of
strong interactions, which are not well understood in dense matter.
The several possible models investigated~\cite{LPMY,physrep}
can be conveniently grouped into three broad categories:
nonrelativistic potential models, field-theoretical models, and
relativistic Dirac-Brueckner-Hartree-Fock models.  In each of these
approaches, the presence of additional softening components such as
hyperons, Bose condensates or quark matter, can be incorporated.

Figure \ref{fig:M-R} displays the mass-radius relation for several
recent EOSs (the abbreviations are explained in Table~\ref{eosname}).
Even a cursory glance indicates that in the mass range from $1-1.5$
M$_\odot$ it is usually the case that the radius has little dependence
upon the mass.  The lone exception is the model GS1, in which a kaon
condensate, leading to considerable softening, appears.  While it is
generally assumed that a stiff EOS leads to both a large maximum mass
and a large radius, many counter examples exist.  For example, MS3 has
a relatively small maximum mass but has large radii compared to most other
EOSs with larger maximum masses.  Also, not
all EOSs with extreme softening have small radii (viz., GS2).
Nonetheless, for stars with mass greater than 1 M$_\odot$, only models
with a large degree of softening can have $R_\infty<12$ km.  Should
the radius of a neutron star ever be accurately determined to satisfy
$R_\infty<12$ km, a strong case can be made for the existence of
extreme softening.

\begin{figure}[h]
\vspace{25pc}
\includegraphics{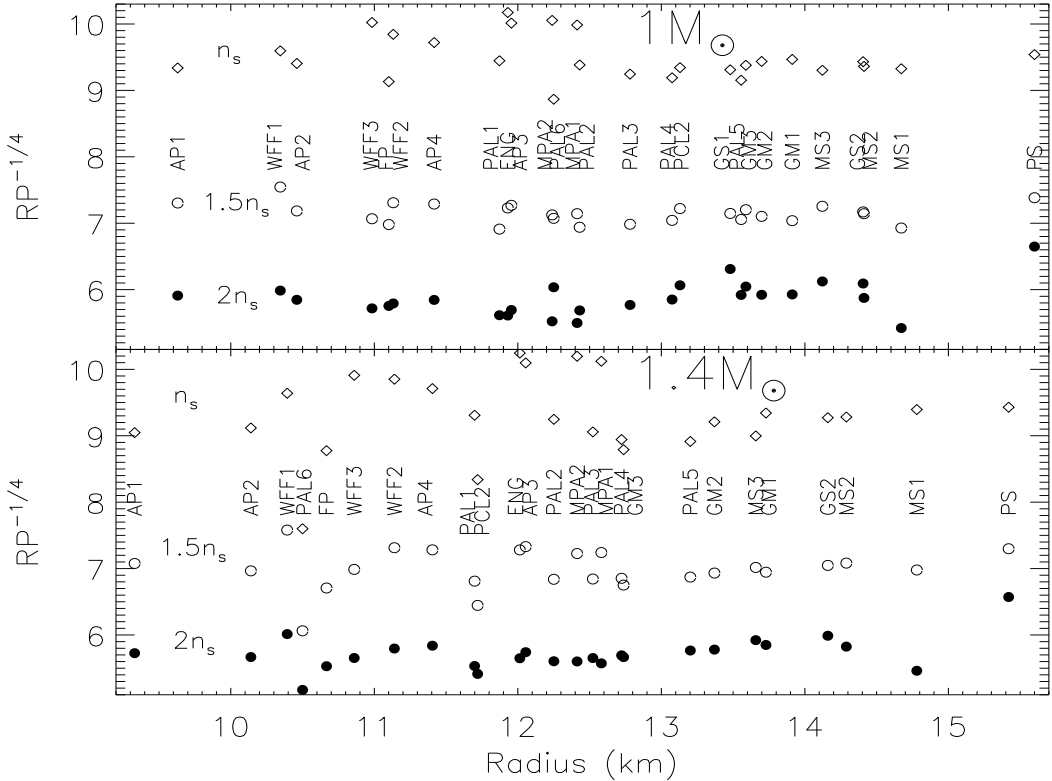}
\caption{Empirical relation between $P$ and $R$ for various EOSs (see
Table~\ref{eosname} for details).  The upper and lower panels show
results for gravitational masses of 1 M$_\odot$ and 1.4 M$_\odot$,
respectively.  Symbols show $PR^{-1/4}$ in units of MeV fm$^{-3}$
km$^{-1/4}$ at the three indicated fiducial densities.}
\label {fig:P-R}
\end{figure}

It is relevant that a Newtonian polytrope with $n=1$ has the property
that the stellar radius is independent of both the mass and central
density.  In fact, numerical relativists have often approximated
equations of state with $n=1$ polytropes.
An $n=1$ polytrope has the property that the radius is proportional to the
square root of the constant $K$ in the polytropic pressure law
$P=K\rho^{1+1/n}$.  This suggests that there might be a quantitative
relation between the radius and the pressure that does not depend upon
the equation of state at the highest densities, which determines the
overall softness or stiffness (and hence, the maximum mass).

To make the relation between matter properties and the nominal neutron star
radius definite, Fig.~\ref{fig:P-R} shows the remarkable empirical
correlation which exists between the radii of 1 and 1.4 M$_\odot$ stars and the
matter's pressure evaluated at densities of 1, 1.5 and 2 $n_s$.
Table~\ref{eosname} explains the EOS symbols used in Fig.~\ref{fig:P-R}.
Despite the relative insensitivity of radius to mass for a particular
``normal'' equation of state, the nominal radius $R_M$, which is defined as the
radius at a particular mass $M$ in solar units, still varies widely with the
EOS employed.  Up to $\sim 5$ km differences are seen in $R_{1.4}$, for
example, in Fig.~\ref{fig:P-R}.  This plot is restricted to EOSs which
have maximum masses larger than about 1.55 M$_\odot$ and to those which do not
have strong phase transitions (such as those due to a Bose condensate or quark
matter).  Such EOSs violate these correlations, especially for the case of 1.4
M$_\odot$ stars.  We emphasize that this correlation is valid only for cold,
catalyzed neutron stars, i.e., it will not be valid for protoneutron stars
which have finite entropies and might contain trapped neutrinos.  The
correlation has the form
\begin{equation}
R \simeq {\rm constant}~\cdot[P(n)]^{0.23-0.26}\,,
\label{correl}
\end{equation}
where $P$ is the total pressure inclusive of leptonic contributions
evaluated at the density $n$.  An exponent of 1/4 was chosen for
display in Fig.~\ref{fig:P-R}, but the correlation holds for a small
range of exponents about this value.  The correlation is marginally
tighter for the baryon density $n=1.5 n_s$ and $2 n_s$ cases.  Thus,
instead of the power 1/2 that the Newtonian polytrope relations would
predict, a power of approximately 1/4 is suggested when the effects of
relativity are included.  The value of the
constant in Eq.~(\ref{correl}) depends upon the chosen density, and
can be obtained from Fig.~\ref{fig:P-R}.

The exponent of 1/4 can be quantitatively understood by using a
relativistic generalization of the $n=1$ polytrope, due to
Buchdahl~\cite{Buchdahl}.  For the EOS
\begin{equation}
\rho=12\sqrt{p_*P}-5P\,,\label{buch}
\end{equation}
where $p_*$ is a constant, there is an analytic solution to Einstein's
equations:
\begin{eqnarray}
e^\nu &\equiv& g_{tt}=(1-2\beta)(1-\beta-u)(1-\beta+u)^{-1}\,;\cr
e^\lambda &\equiv&
g_{rr}=(1-2\beta)(1-\beta+u)(1-\beta-u)^{-1}(1-\beta+\beta\cos Ar^\prime)^{-2}\,;\cr
8\pi PG/c^4 &=& A^2u^2(1-2\beta)(1-\beta+u)^{-2}\,;\cr
8\pi\rho G/c^2&=& 2A^2u(1-2\beta)(1-\beta-3u/2)(1-\beta+u)^{-2}\,;\cr
u&=&\beta(Ar^\prime)^{-1}\sin Ar^\prime\,;\qquad
r = r^\prime(1-\beta+u)(1-2\beta)^{-1}\,;\cr
A^2 &=& 288\pi p_*Gc^{-4}(1-2\beta)^{-1};\qquad R=\pi(1-\beta)(1-2\beta)^{-1}A^{-1}.
\label{buch1}
\end{eqnarray}

The free parameters of this solution are $\beta\equiv GM/Rc^2$ and the
scale $p_*$.  Note that $R\propto p_*^{-1/2}(1+\beta^2/2+\dots)$, so
for a given value of $p_*$, the radius increases only very slowly
with mass, exactly as expected from an $n=1$
Newtonian polytrope.  It is instructive to analyze the response of $R$
to a change of pressure at some fiducial density $\rho$, for a fixed
mass $M$.  One finds
\begin{equation}
{d\ln R\over d\ln P}\Biggr|_{\rho,M} = {{d\ln R\over d\ln
p_*}\Bigr|_\beta {d\ln p_*\over d\ln P}\Bigr|_{\rho}\over1+{d\ln
R\over d\ln\beta}\Bigr|_{p_*}}=
\Biggl(1-{5\over6}\sqrt{P\over p_*}\Biggr){(1-\beta)(1-2\beta)\over2(1-3\beta+3\beta^2)}.
\end{equation}
In the limit $\beta\rightarrow0, P\rightarrow0$ and $d\ln
R/d\ln P\rightarrow1/2$, the value characteristic of an $n=1$ Newtonian
polytrope. Finite values of $\beta$ and $P$ render the
exponent smaller than 1/2.  If the stellar radius is about 15 km,
$p_*=\pi/(288 R^2)\approx4.85\cdot10^{-5}$ km$^{-2}$.  If the fiducial
density is $\rho\approx 1.5m_bn_s\approx2.02\cdot10^{-4}$ km$^{-2}$
(with $m_b$ the baryon mass), Eq.~(\ref{buch}) implies that
$P\approx8.5\cdot10^{-6}$ km$^{-2}$.  For $M=1.4$ M$_\odot$, the value
of $\beta$ is 0.14, and $d\ln R/d\ln P\simeq0.31$.  This result is
mildly sensitive to the choices for $\rho$ and $R$, and the Buchdahl
solution is not a perfect representation of realistic EOSs;
nevertheless, it provides a reasonable explanation of the
correlation in Eq.~(\ref{correl}).

The existence of this correlation is significant because, in large
part, the pressure of degenerate matter near the nuclear saturation
density $n_s$ is determined by the symmetry properties of the EOS.
Thus, the measurement of a neutron star radius, if not so small as to
indicate extreme softening, could provide an important clue to the
symmetry properties of matter.  In either case, valuable information
is obtained.

\begin{figure}[h]
\vspace{23pc}
\includegraphics{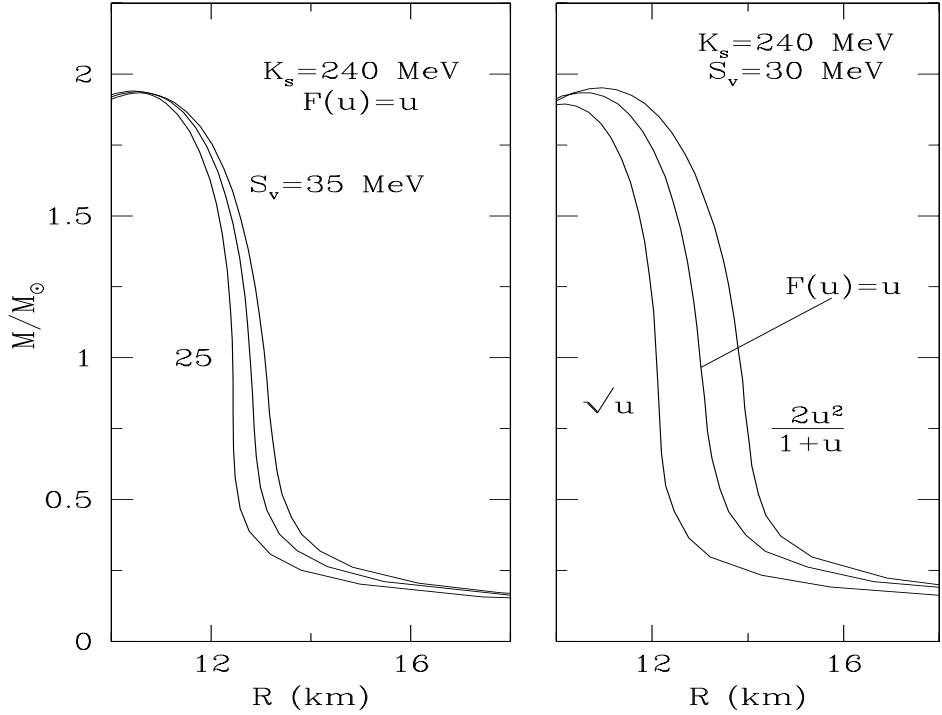}
\caption{Left panel: $M-R$ curves for selected PAL parametrizations~\cite{PAL}
 showing the
sensitivity to symmetry energy.  The left panel shows variations
arising from different choices of the symmetry energy at the nuclear
saturation density $S_v=S_v(n_s)$; the right panel shows variations
arising from different choices of the density dependence of the
potential part of the symmetry energy $F(u)=S_v(n)/S_v(n_s)$ where $u=n/n_s$.}
\label{fig:alp9}
\end{figure}

The specific energy of nuclear matter near the saturation density may
be expressed as an expansion in the asymmetry $(1-2x)$, as
displayed in Eq.~(\ref{bulk}), that can be terminated after the quadratic
term~\cite{PAL}.  Leptonic contributions must be added to
Eq.~(\ref{bulk}) to obtain the total energy and pressure; the electron
energy per baryon is $f_e=(3/4)\hbar cx(3\pi^2nx)^{1/3}$.  Matter in
neutron stars is in beta equilibrium, i.e., $\mu_e - \mu_n + \mu_p =
\partial (f_{bulk}+f_e)/\partial x=0$, so the electronic contributions
may be eliminated to recast the pressure as~\cite{Ppuri}
\begin{eqnarray}
P=n^2\Biggl[S_v^\prime(n)(1-2x)^2+{xS_v(n)\over n}(1-2x)+\cr
{K_s\over9n_s}\Bigl({n\over
n_s}-1\Bigr)-{K_s^\prime\over54n_s}\Bigl({n\over n_s}-1\Bigr)^2\Biggr]\,,
\end{eqnarray}
where $x$ is now the beta equilibrium value.  At the saturation density,
\begin{eqnarray}
P_s=n_s(1-2x_s)[n_sS_v^\prime(n_s)(1-2x_s)+S_v x_s]\,,
\end{eqnarray}
where the equilibrium proton fraction at $n_s$ is
\begin{eqnarray}
x_s\simeq(3\pi^2 n_s)^{-1}(4S_v/\hbar c)^3 \simeq 0.04
\end{eqnarray}
for $S_v=30$ MeV. Due to the small value of $x_s$, one finds that
$P_s\simeq n_s^2 S_v^\prime(n_s)$.  If the pressure is evaluated at a
larger density, other nuclear parameters besides $S_v$ and
$S_v^\prime(n_s)$, become significant.  For $n=2n_s$, one thus has
\begin{eqnarray}
P(2n_s)\simeq 4n_s [n_sS_v^\prime(2n_s)+(K_s - K_s^\prime/6)/9] \,.
\end{eqnarray}
If it is assumed that $S_v(n)$ is linear in density, $K_s\sim220$ MeV and
$K_s^\prime\sim2000$ MeV (as indicated in Eq.~\ref{comp}), the symmetry
contribution is still about 70\% of the total.

The sensitivity of the radius to the symmetry energy is graphically
shown by the parametrized EOS of PAL~\cite{PAL} in
Fig.~\ref{fig:alp9}.  The symmetry energy function $S_v(n)$ is a
direct input in this parametrization.  The figure shows the dependence
of mass-radius trajectories as the quantities $S_v$ and $S_v(n)$ are
alternately varied.  Clearly, the density dependence of $S_v(n)$ is
more important in determining the neutron star radius.  Note also the
weak sensitivity of the maximum neutron star mass to $S_v$.

At present, experimental guidance concerning the density dependence of the
symmetry energy is limited and mostly based upon the division of the nuclear
symmetry energy between volume and surface contributions, as discussed in the
previous section.  Upcoming experiments involving heavy-ion collisions (at GSI,
Darmstadt), which might sample densities up to $\sim (3-4)n_s$, will be limited
to analyzing properties of the symmetric nuclear matter EOS through a study of
matter, momentum, and energy flow of nucleons.  Thus, studies of heavy nuclei
far off the neutron drip lines will be necessary in order to pin down the
properties of the neutron-rich regimes encountered in neutron stars.

\subsection{Neutron Star Moments of Inertia and Binding Energies}

Besides the stellar radius, other global attributes of
neutron stars are potentially observable, including the moment of inertia
and the binding energy.  These quantities depend
primarily upon the ratio $M/R$ as opposed to details of the EOS,
as can be readily seen by evaluating them using analytic
solutions to Einstein's equations.  Although over 100 analytic
solutions to Einstein's equations are known~\cite{Delgaty}, nearly all of
them are physically unrealistic.  However, three analytic solutions are
of particular interest in neutron star structure.

The first is the well-known Schwarzschild interior solution for an
incompressible fluid, $\rho=\rho_c$, where $\rho$ is the mass-energy
density.  This is mostly of interest because it determines the maximum
compression $\beta=GM/Rc^2$ for a neutron star, namely 4/9, based upon
the pressure being finite.  Two aspects of the incompressible fluid
that are physically unrealistic, however, include the fact that the
sound speed is everywhere infinite, and that the density does not
vanish on the star's surface.

The second analytic solution, B1, due to Buchdahl~\cite{Buchdahl}, is
described in Eq.~(\ref{buch1}).

\begin{figure}[h]
\vspace{28pc}
\includegraphics{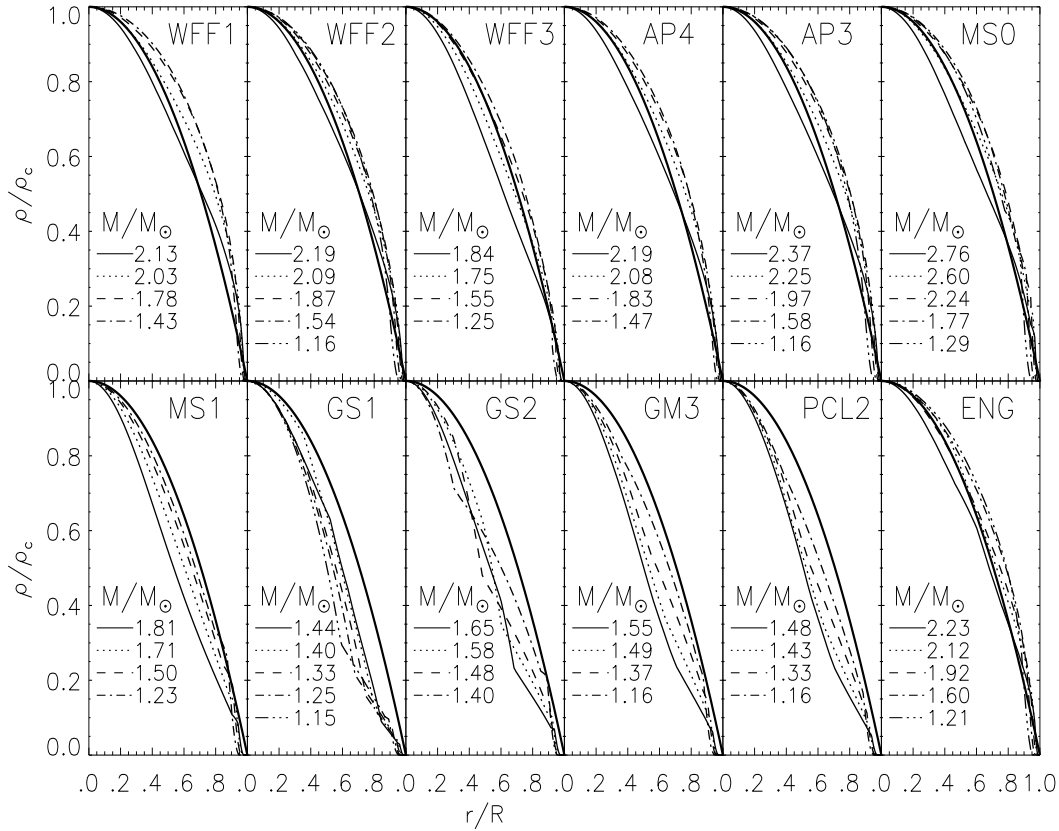}
\caption{Each panel shows mass-energy density profiles in the
interiors of selected stars (masses indicated) ranging from about 1.2
M$_\odot$ to the maximum mass (solid line) for
the given equation of state (see
Table~\ref{eosname}).  The thick black lines show the simple quadratic
approximation $1-(r/R)^2$.}
\label{fig:prof}
\end{figure}

The third analytic solution (TolVII) was discovered by Tolman~\cite{Tolman} in
1939, and is the case when the mass-energy density
$\rho$ varies quadratically, that is,
\begin{equation}
\rho=\rho_c[1-(r/R)^2].
\end{equation}
In fact, this is an adequate representation, as displayed in
Fig.~\ref{fig:prof} for neutron stars more massive than 1.2 M$_\odot$.  The
equations of state used are listed in Table~\ref{eosname}.  The largest
deviations from this general relation exist for models with extreme softening
(GS1, GS2, PCL2) and which have relatively low maximum masses (see
Fig.~\ref{fig:M-R}).  It is significant that all models must, of course,
approach this behavior at both extremes $r\rightarrow0$ and $r\rightarrow R$.

Because the Tolman solution is often overlooked in the
literature (for exceptions, see, for example, Refs.~\cite{Delgaty,Indians})
it is
summarized here.  It is useful in establishing interesting and simple relations
that are insensitive to the equation of state.  In terms of the variable
$x=r^2/R^2$ and the parameter $\beta$, the assumption
$\rho=\rho_c(1-x)$ results in $\rho_c=15\beta c^2/(8\pi GR^2)$.  The
solution of Einstein's equations for this density distribution is:
\begin{eqnarray}
e^{-\lambda} &=& 1-\beta x(5-3x)\,,\qquad e^\nu = (1-5\beta/3)\cos^2\phi\,,
\cr
P &=& {c^4\over4\pi R^2 G}[\sqrt{3\beta e^{-\lambda}}\tan\phi-{\beta\over2}(5-3x)\,,
\qquad n= {\rho c^2+P\over m_bc^2}{\cos\phi\over\cos\zeta}\,, \cr
\phi &=& (w_1-w)/2+\zeta\,, \quad \phi_c = \phi(x=0)\,, \quad \zeta = \tan^{-1}\sqrt{\beta/[3(1-2\beta)]}\,,\cr
w &=& \log[x-5/6+\sqrt{e^{-\lambda}/(3\beta)}]\,, \qquad w_1 = w(x=1)\,.
\end{eqnarray}
The central values of $P/\rho c^2$ and $c_s^2$ are
\begin{equation}
{P\over\rho c^2}\Biggr|_c={2\over15}\sqrt{3\over\beta}\Bigr({c_{sc}\over
c}\Bigr)^2\,,\quad \Bigr({c_{sc}\over c}\Bigr)^2=\tan\phi_c\Bigr(\tan\phi_c+\sqrt{\beta\over3}\Bigr)\,.
\end{equation}
This solution, like that of Buchdahl's, is scale-free, with the
parameters $\beta$ and $\rho_c$ (or $M$ or $R$).  Here, $n$ is the baryon
density, $m_b$ is the nucleon mass, and $c_{sc}$ is the sound speed at
the star's center.  When $\phi_0=\pi/2$, or $\beta\approx0.3862$,
$P_c$ becomes infinite, and when $\beta\approx0.2698$, $c_{sc}$
becomes causal ({i.e., $c$).  Recall that for an incompressible fluid,
$P_c$ becomes infinite when $\beta=4/9$.  For the Buchdahl solution,
$P_c$ becomes infinite when $\beta=2/5$ and the causal limit is
reached when $\beta=1/6$.  For comparison, if causality is enforced at
high densities, it has been empirically determined that
$\beta<0.34$~\cite{LPMY,glen}.

The general applicability of these exact solutions can be gauged by analyzing
the moment of inertia, which, for a star uniformly
rotating with angular velocity $\Omega$, is
\begin{equation}I=(8\pi/3)\int_0^R r^4(\rho+P/c^2)e^{(\lambda-\nu)/2}
(\omega/\Omega) dr\,.\label{inertia}\end{equation}
The metric function $\omega$ is a solution of the equation
\begin{equation}
d[r^4e^{-(\lambda+\nu)/2}\omega^\prime]/dr+4r^3\omega
de^{-(\lambda+\nu)/2}/dr=0
\label{diffomeg}
\end{equation}
with the surface boundary condition
\begin{equation}\omega_R=\Omega-{R\over3}\omega^\prime_R
=\Omega\left[1-{2GI\over R^3c^2}\right].
\label{boundary}
\end{equation}
The second equality in the above follows from the definition of $I$ and the TOV
equation.  Writing $j=\exp[-(\nu+\lambda)/2]$, the
TOV equation becomes
\begin{equation}
j^\prime=-4\pi Gr(P/c^2+\rho)je^\lambda/c^2\,.
\end{equation}
Then, one has
\begin{equation}
I=-{2c^2\over3G}\int {\omega\over\Omega}r^3dj =
{c^2R^4\omega^\prime_R\over6G\Omega} \,. \end{equation}

\begin{figure}[h]
\vspace{23pc}
\includegraphics{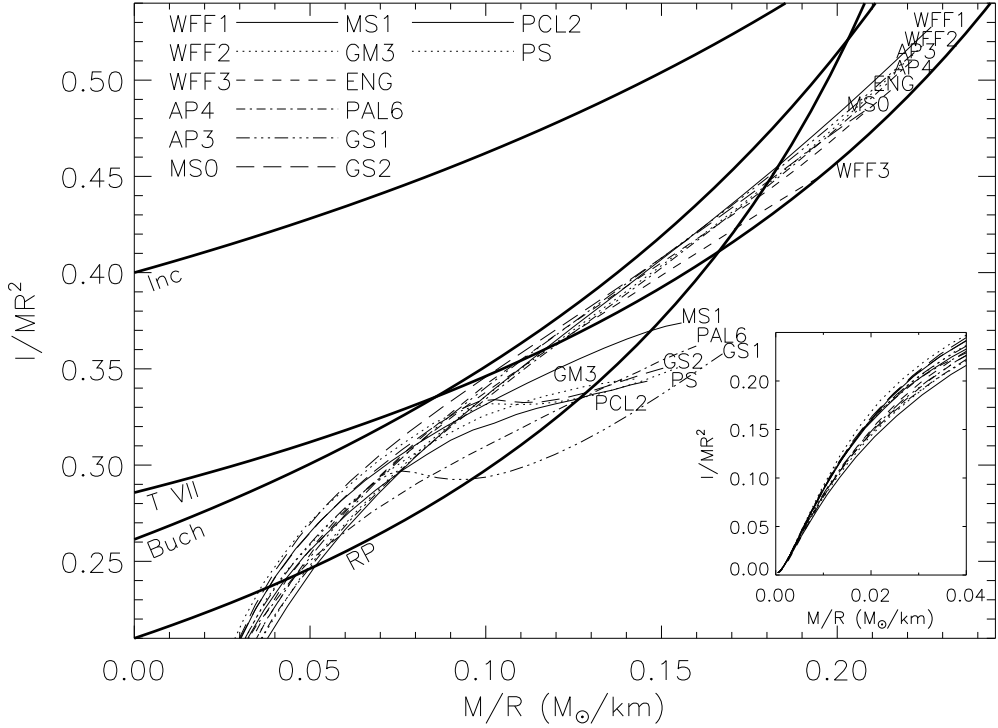}
\caption{The moment of inertia $I$ in units of $MR^2$ for the equations of
state listed in Table~\ref{eosname}.  $I_{Inc}, I_{B 1}, I_{VII}$ and $I_{RP}$
are approximations described in the text.}
\label{mominert}
\end{figure}

Unfortunately, an
analytic representation of $\omega$ or the moment of inertia for any of the
three exact solutions is not available.  However, approximations which are
valid to within 0.5\% are
\begin{eqnarray}
I_{Inc}/MR^2 &\simeq&  2(1-0.87\beta-0.3\beta^2)^{-1}/5\,, \\
I_{B1}/MR^2 &\simeq& (2/3-4/\pi^2)(1-1.81\beta+0.47\beta^2)^{-1}\,, \\
I_{T VII}/MR^2 &\simeq& 2(1-1.1\beta-0.6\beta^2)^{-1}/7\,.
\end{eqnarray}
In each case, the small $\beta$ limit reduces to the corresponding Newtonian
results.  Fig.~\ref{mominert} indicates
that the Tolman approximation is rather good.  Ravenhall
\& Pethick~\cite{RP} suggested that the expression
\begin{equation}
I_{RP}/MR^2\simeq0.21/(1-2u)
\end{equation}
was a good approximation for the moment of inertia; however, we find
that this expression is not a good overall fit, as shown in
Fig.~\ref{mominert}.  For low-mass stars ($\beta<0.12$), none of these
approximations is suitable, but it is unlikely that any neutron stars
are this rarefied.  It should be noted that the Tolman approximation
does not adequately approximate some EOSs, especially ones that are
relatively soft, such as GM3, GS1, GS2, PAL6 and PCL2.

The binding energy formally represents the energy gained by assembling
$N$ baryons.  If the baryon mass is $m_b$, the binding energy is
simply $BE=Nm_b-M$ in mass units.  However, the quantity $m_b$ has various
interpretations in the literature.  Some authors assume it is about
940 MeV/$c^2$, the same as the neutron or proton mass.  Others assume
it is about 930 MeV/$c^2$, corresponding to the mass of C$^{12}$/12 or
Fe$^{56}$/56.  The latter would yield the energy released in a
supernova explosion, which consists of the energy released by the
collapse of a white-dwarf-like iron core, which itself is considerably
bound.  The difference, 10 MeV per baryon, corresponds to a shift of
$10/940\simeq0.01$ in the value of $BE/M$.  In any case, the binding
energy is directly observable from the detection of neutrinos from a
supernova event; indeed, it would be the most precisely determined
aspect.

\begin{figure}[h]
\vspace{25pc}
\includegraphics{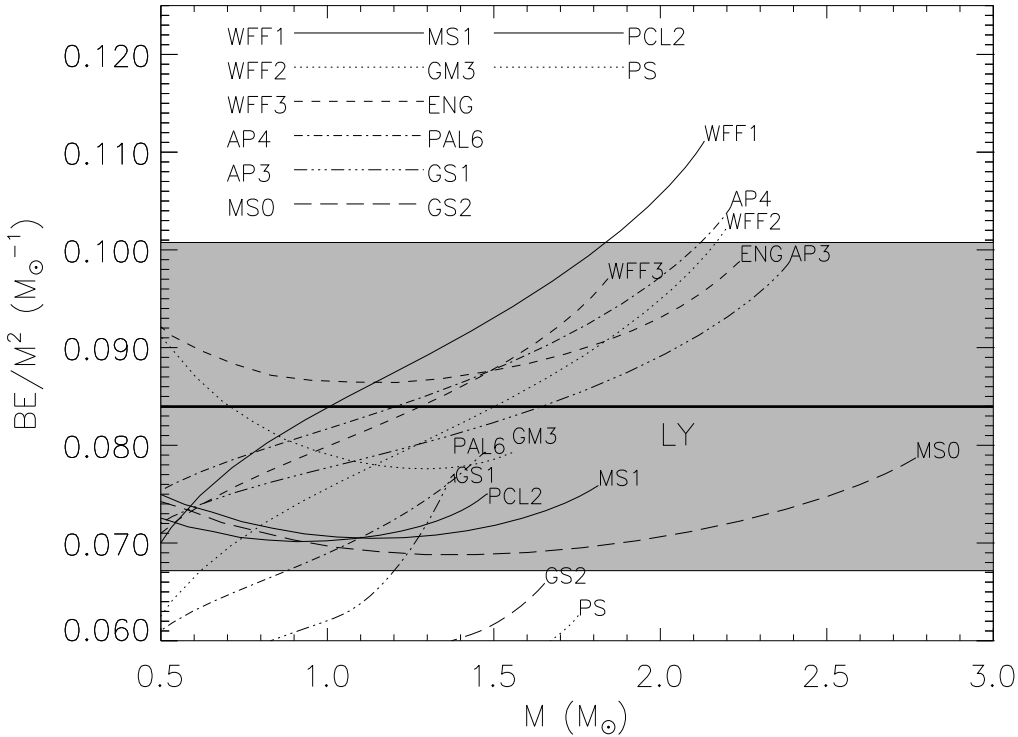}
\caption{The binding energy of neutron stars as a function of stellar mass for
the equations of state listed in Table~\ref{eosname}.  The predictions of
Eq.~(\ref{lybind}) are shown by the shaded region.}
\label{bind}
\end{figure}

Lattimer \& Yahil~\cite{LY} suggested that the binding energy could be
approximated as
\begin{equation}
BE\approx 1.5\cdot10^{51} (M/{\rm M}_\odot)^2 {\rm~ergs} = 0.084
(M/{\rm M}_\odot)^2 {\rm~M}_\odot\,.
\label{lybind}
\end{equation}
This formula, in general, is accurate to about $\pm20$\%.  The largest
deviations are for extremely soft EOSs,  as shown in
Fig.~\ref{bind}.

\begin{figure}[h]
\vspace{25pc}
\includegraphics{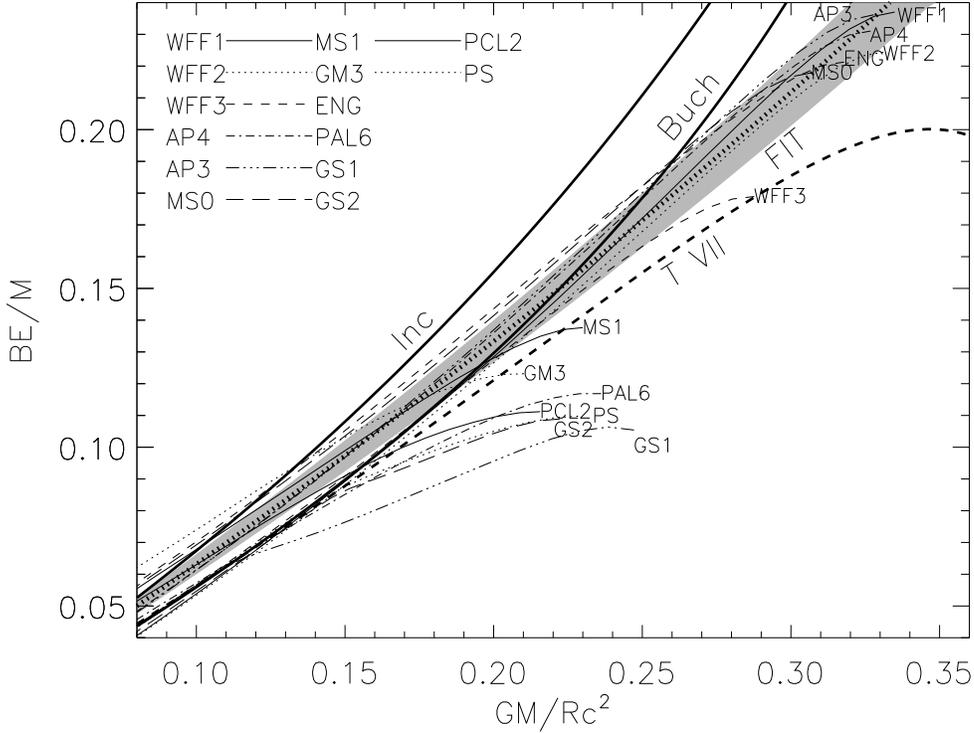}
\caption{The binding energy per unit gravitational mass as a function of
compactness for the equations of state listed in Table~\ref{eosname}.  The
shaded region shows the prediction of Eq.~(\ref{newbind})
with $\pm5$\% errors.}
\label{bind1}
\end{figure}

However, a more accurate representation of the binding energy is given by
\begin{equation}
BE/M \simeq 0.6\beta/(1-0.5\beta)\,, \label{newbind}
\end{equation}
which incorporates some radius dependence.  Thus, the observation of supernova
neutrinos, and the estimate of the total radiated neutrino energy, will yield
more accurate information about $M/R$ than about $M$ alone.

In the cases of the incompressible fluid and the Buchdahl solution, analytic
results for the binding energy can be found:
\begin{eqnarray}
BE_{Inc}/M &=& {3\over4\beta}\Bigl({\sin^{-1}\sqrt{2\beta}\over
\sqrt{2\beta}}-\sqrt{1-2\beta}\Bigr)-1\,, \\
BE_{B1}/M &=&  (1-1.5\beta)\sqrt{1-2\beta}(1-\beta)^{-1}-1\,.\label{analbind}
\end{eqnarray}
The analytic results, the Tolman VII solution, and the fit of
Eq.~(\ref{newbind}) are compared to some recent equations of state in
Fig.~\ref{bind1}.  It can be seen that, except for very soft cases
like PS, PCL2, PAL6, GS1 and GS2, both the Tolman VII and Buchdahl
solutions are rather realistic.

\subsection{Crustal Fraction of the Moment of Inertia}

In the investigation of pulsar glitches, many models associate the
glitch size with the fraction of the moment of inertia which resides
in the star's crust, usually defined to be the region in which dripped
neutrons coexist with nuclei.  The high-density crust boundary is set
by the phase boundary between nuclei and uniform matter, where the
pressure is $P_t$ and the density $n_t$.  The low-density boundary is
the neutron drip density, or for all practical purposes, simply the
star's surface since the amount of mass between the neutron drip point
and the surface is negligible.  We define $\Delta R$ to be the
distance between the points where the density is $n_t$ and zero.  One
can apply Eq.~(\ref{inertia}) to determine the moment of inertia of the
crust alone with the assumptions that $P/c^2<<\rho$, $m(r)\simeq M$,
and $\omega j\simeq\omega_R$ in the crust.  One finds
\begin{equation}
\Delta I\simeq{8\pi\over3}{\omega_R\over\Omega}\int_{R-\Delta
R}^R \rho r^4e^\lambda dr\simeq
{8\pi\over3GM}{\omega_R\over\Omega}\int_0^{P_t}r^6dP\,,
\label{deltai}
\end{equation}
where $M$ is the star's total mass and the TOV equation was used in
the last step.  In the crust, the fact that the EOS is
of the approximate polytropic form $P\simeq K\rho^{4/3}$ can be used
to find an approximation for the integral $\int r^6dP$, {\em viz.}
\begin{equation}
\int_0^{P_t}r^6dP\simeq P_tR^6\left[1+
{2P_t\over n_t m_nc^2}{(1+7\beta)(1-2\beta)\over\beta^2}\right]^{-1}\,.
\end{equation}
Since the approximation Eq.~(\ref{newbind}) gives $I$ in terms of $M$ and $R$,
and $\omega_R/\Omega$ is given in terms of $I$ and $R$ in Eq.~(\ref{boundary}),
the quantity $\Delta I/I$ can thus be expressed as a function of $M$ and $R$
with the only dependence upon the equation of state (EOS) arising from the
values of $P_t$ and $n_t$; there is no explicit dependence upon the
higher-density EOS.  However, the major dependence is upon the value of $P_t$,
since $n_t$ enters only as a correction.  We then find
\begin{equation}{\Delta I\over I}\simeq{28\pi P_t
R^3\over3 Mc^2}{(1-1.67\beta-0.6\beta^2)\over\beta}\left[1+{2P_t\over n_t
m_bc^2}{(1+7\beta)(1-2\beta)\over\beta^2}\right]^{-1}.
\label{dii}
\end{equation}

In general, the EOS parameter $P_t$, in the units of MeV fm$^{-3}$, varies over
the range $0.25<P_t<0.65$ for realistic EOSs.  The determination
of this parameter requires a calculation of the structure of matter containing
nuclei just below nuclear matter density that is consistent with the assumed
nuclear matter EOS.  Unfortunately, few such calculations have been
performed.
Like the fiducial pressure at and above nuclear density which appears in the
relation Eq.~(\ref{correl}), $P_t$ should depend sensitively
 upon the behavior of the
symmetry energy near nuclear density.

Choosing $n_t=0.07$ fm$^{-3}$, we compare Eq.~(\ref{dii}) in Fig.~\ref{fig:M-R}
with full structural calculations.  The agreement is good.  We also note that
Ravenhall \& Pethick~\cite{RP} developed a different, but nearly equivalent,
formula for the quantity $\Delta I/I$ as a function of $M, R, P_t$ and $\mu_t$,
where $\mu_t$ is the neutron chemical potential at the core-crust phase
boundary.  This prediction is also displayed in Fig.~\ref{fig:M-R}.

Link, Epstein \& Lattimer~\cite{Link} established a lower limit to the radii of
neutron stars by using a constraint derived from pulsar glitches.  They showed
that glitches represent a self-regulating instability for which the star
prepares over a waiting time.  The angular momentum requirements of glitches in
the Vela pulsar indicate that more than
$0.014$ of the star's moment of inertia drives
these events.  If glitches originate in the liquid of the inner crust, this
means that $\Delta I/I>0.014$.  A minimum radius can be found by combining this
constraint with the largest realistic value of $P_t$ from any equation of
state.  Stellar models that are compatible with this constraint must fall to
the right of the $P_t=0.65$ MeV fm$^{-3}$ contour in Fig.~\ref{fig:M-R}.  This
imposes a constraint upon the radius, namely that
$R>3.6+3.9 M/{\rm M}_\odot$ km.

\section{The Merger of a Neutron Star with a Low-Mass Black Hole}
The general problem of the origin and evolution of systems containing a neutron
star and a black hole was first detailed by Lattimer \& Schramm~\cite{LSch},
although the original motivation was due to Schramm.  Although speculative
at the time,
Schramm insisted that this would prove to be an interesting topic from the
points of view of nucleosynthesis and gamma-ray emission.  The contemporaneous
discovery~\cite{HT} of the first-known binary system containing twin compact
objects, PSR 1913+16, which was also found to have an orbit which would decay
because of gravitational radiation within $10^{10}$ yr, bolstered his argument.
Eventually, this topic formed the core of Lattimer's thesis~\cite{thesisL},
and the
recent spate of activity, a quarter century later, in the investigation of the
evolution and mergers of such compact systems has wonderfully demonstrated
Schramm's prescience.

Compact binaries form naturally as the result of evolution of massive
stellar binaries.  The estimated lower mass limit for supernovae (and
neutron star or black hole production) is approximately 8 M$_\odot$.
Observationally, the number of binaries formed within a given
logarithmic separation is approximately constant, but the relative
mass distributions are uncertain.  There is some indication that the
distribution in binary mass ratios might be flat.  The number of
possible progenitor systems can then be estimated.  Most progenitor systems
do not survive the more massive star becoming a supernova. In the
absence of a kick velocity it is easily found that the loss of more
than half of the mass from the system will unbind it.  However, the
fact that pulsars are observed to have mean velocites in excess of a
few hundred km/s implies that neutron stars are usually produced with
large ``kick'' velocities originating in the supernova explosion.  In
the case that the kick velocity, which is thought to be randomly
directed, opposes the star's orbital velocity, the chances
of the post-supernova binary remaining intact increases.  In
addition, the separation in a surviving binary will be reduced
significantly.  Subsequent evolution then progresses to the supernova
explosion of the companion.  More of these systems survive because in
many cases the more massive component explodes.  But the surviving
systems should both have greatly reduced separations and orbits with
high eccentricity.

\begin{figure}[h]
\vspace{24pc}
\includegraphics{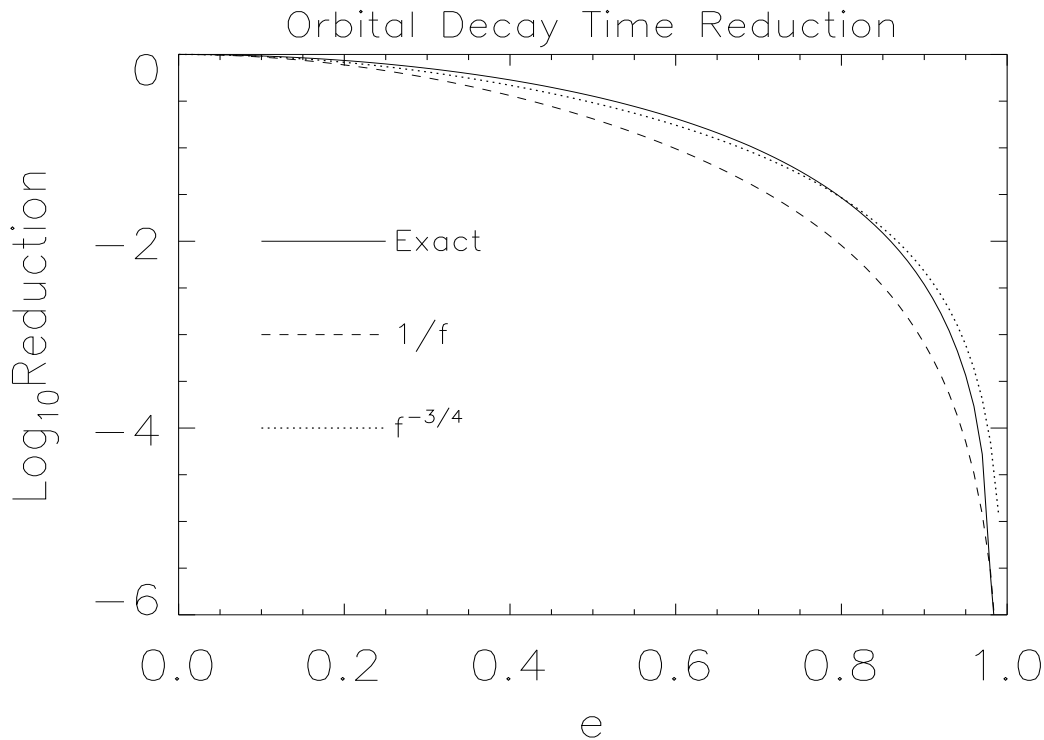}
\caption{The reduction of the gravitational radiation orbital decay time as a
function of initial orbital eccentricity.  The dashed line is the inverse of
the Peters~\cite{Peters} $f$ function; the dotted line shows $f^{-3/4}$,
which reasonably reproduces the exact result.}
\label{reduc}
\end{figure}

Gravitational radiation then causes the binary's
orbit to decay, such that circular orbits of two masses $M_1$ and
$M_2$ with initial semimajor axes $a$
satisfying
\begin{equation}
a<2.8[M_1M_2(M_1+M_2)/{\rm M}_\odot^3]^{1/4} {\rm R}_\odot\,,
\label{decay}
\end{equation}
will fully decay within
the age of the Universe ($\sim10^{10}$ yr).  Highly
eccentric orbits will decay much faster, as shown in Fig.~\ref{reduc}.
The dashed curve shows the inverse of the factor~\cite{Peters} by which
the gravitational wave luminosity of an eccentric system exceeds that
of a circular system:
\begin{equation}
f=(1+73e^2/24+37e^4/96)(1-e^2)^{-7/2}.
\label{peters}
\end{equation}
Because the eccentricity also decays, the exact reduction factor is
not as strong as $1/f$.  A
reasonable approximation to the exact result is $f^{-3/4}$, shown by the
dotted line in Fig.~\ref{reduc}.  The coefficient 2.8 in
Eq.~(\ref{decay}) is increased by a factor of $f^{-3/16}$ or about 2 for
moderate eccentricities.

Ref.~\cite{LSch} argued that mergers of neutron stars and black holes, and the
subsequent ejection of a few percent of the neutron star's mass, could easily
account for all the {\em r}-process nuclei in the cosmos.  Ref.~\cite{LSch} is
also the earliest reference to the idea that compact object binary mergers are
associated with gamma-ray bursts.  A later seminal contribution by Eichler,
Livio, Piran \& Schramm~\cite{eichler} argued that mergers of neutron stars
occur frequently enough to explain the origin of gamma-ray bursters.

Since the timescale of gamma-ray bursts, being of order seconds to several
minutes, is much longer than the coalescence timescale of a binary merger
(which is of order the orbital frequency at the last stable orbit, a few
milliseconds), it is believed that a coalescence involves the formation of an
accretion disc.  Although neutrino emission from accreting material, resulting
in neutrino-antineutrino annihilation along the rotational axis, has been
proposed as a source of gamma rays, it seems more likely that amplification of
magnetic fields within the disc might trigger observed bursts.  In either case,
the lifetime of the accretion disc is still problematic, if it is formed by the
breakup of the neutron star near the Roche limit.  Its lifetime would probably
be only about a hundred times greater than the orbital frequency, or less than
a second.  However, this timescale would be considerably enhanced if the
accretion disc could be formed at larger radii than the Roche limit.
A possible mechanism is stable mass transfer from the neutron star to
the black hole that would cause the neutron star to spiral away as it
loses mass~\cite{Kochanek,PZ}.

The classical Roche limit is based upon an incompressible fluid of density
$\rho$ and mass $M_2$ in orbit about a mass $M_1$.  In Newtonian gravity, this
limit is
\begin{equation}
R_{Roche, Newt}=(M_1/0.0901\pi\rho)^{1/3}= 19.2
(M_1/{\rm M}_\odot
\rho_{15})^{1/3}{\rm~km}\,,\label{roche}
\end{equation}
where $\rho_{15}=\rho/10^{15}$ g cm$^{-3}$.  Using general relativity,
Fishbone~\cite{Fishbone} found that at the last stable circular orbit
(including the case when the black hole is rotating) the number 0.0901
in Eq.~(\ref{roche}) becomes 0.0664.  In geometrized units,
$R_{Roche}/M_1=13(14.4)(M_1^2\rho_{15}/{\rm M}_\odot^2)^{-1/3}$, where
the numerical coefficient refers to the Newtonian (last stable orbit
in GR) case.  In other words, if the neutron star's mean density is
$\rho_{15}=1$, the Roche limit is encountered beyond the last stable
orbit if the black hole mass is less than about 5.9 M$_\odot$.  Thus,
for small enough black holes, mass overflow and transfer from the
neutron star to the black hole could begin outside the last stable
circular orbit.  And, as now discussed, the mass transfer
may proceed stably for some
considerable time.  In fact, the neutron star might move to 2--3 times
the orbital radius where mass transfer began.  This would provide a
natural way to lengthen the lifetime of an accretion disc, by simply
increasing its size.

The final evolution of a compact binary is now discussed.  Define
$q=m_{ns}/M_{BH}$, $\mu=m_{ns}M_{BH}/M$, and $M=M_{BH}+m_{ns}$, where
$m_{ns}$ and $M_{BH}$ are the neutron star and black hole masses,
respectively. The orbital angular momentum is
\begin{equation}
J^2=G\mu^2 Ma=GM^3aq^2/(1+q)^4\,.
\label{j2}
\end{equation}
We can employ Paczy\'nski's~\cite{pacz} formula for the Roche radius of the
secondary:
\begin{equation}
R_\ell/a=0.46[q/(1+q)]^{1/3}\,,
\label{rl}
\end{equation}
or a better fit by Eggleton~\cite{eggleton}:
\begin{equation}
R_\ell/a=0.49[.6+q^{-2/3}\ln(1+q^{1/3})]^{-1}\,.
\label{eggleton}
\end{equation}

\begin{figure}[h]
\vspace{22pc}
\includegraphics{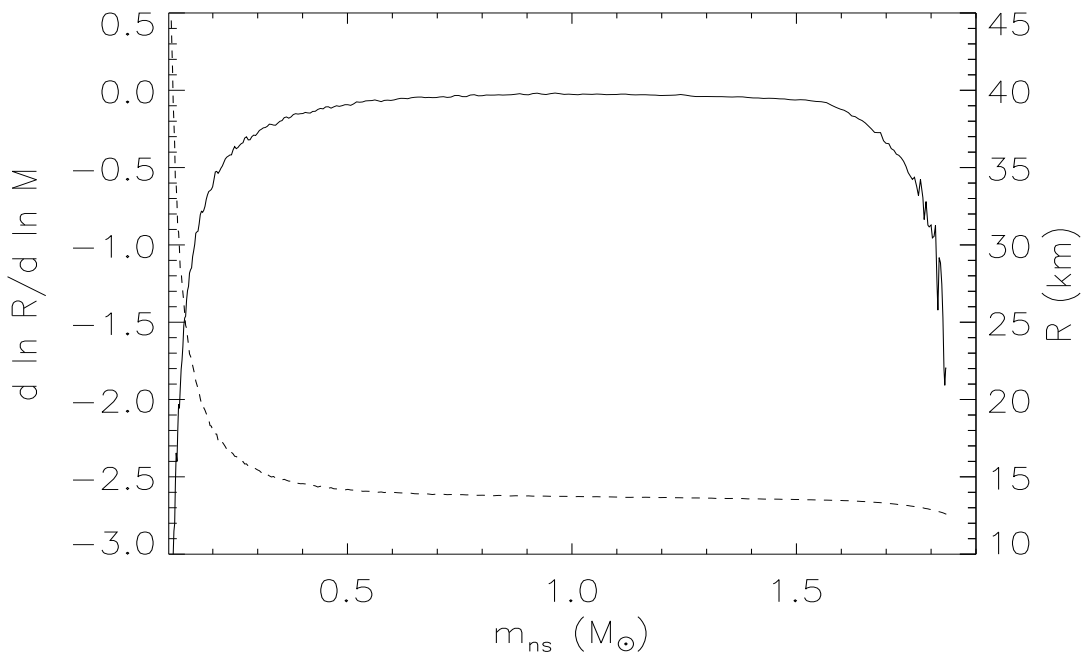}
\caption{$d\ln R/d\ln m_{ns}\equiv\alpha$ (solid curve) and neutron star radius
$R$ (dashed curve) as functions of neutron star mass $m_{ns}$ for a typical
dense matter equation of state.}
\label{alpha}
\end{figure}
The orbital separation $a$ at the moment of mass transfer is obtained by
setting $R_\ell=R$, the neutron star radius.  For stable mass transfer, the
star's radius has to increase more quickly than the Roche radius as mass is
transferred.  Thus, we must have, using Paczy\'nski's formula,
\begin{equation}
{d\ln R\over d\ln m_{ns}}\equiv\alpha\ge{d\ln R_\ell\over d\ln m_{ns}}= {d\ln
a\over d\ln m_{ns}}+{1\over3}\label{dlnrdlnm2}\end{equation}
for stable mass transfer.
$\alpha$ is defined in this expression, and is shown
in Fig.~\ref{alpha} for a typical EOS.
If the mass
transfer is conservative, than $\dot J=\dot J_{GW}$, where
\begin{equation}
\dot J_{GW}=-{32\over5}{G^{7/2}\over c^5}{\mu^2 M^{5/2}\over a^{7/2}}=
-{32\over5}{G^{7/2}\over c^5}{q^2 M^{9/2}\over(1+q)^4a^{7/2}}\label{dotgw}
\end{equation}
and
\begin{equation}
{\dot J\over J}={\dot a\over2a}+{\dot q(1-q)\over
q(1+q)}\,.\label{dotj}
\end{equation}
This leads to
\begin{equation}
\dot q\left({\alpha\over2}+{5\over6}-q\right)\ge -{32\over5}{G^3\over c^5}{q^2
M^3\over(1+q)a^4}\,.\label{dotq}\end{equation} Since $m_{ns}<M_{BH}$,
$\dot q\le0$, and the condition for stable
mass transfer is simply $q\le5/6+\alpha/2$.  For moderate
mass neutron stars, $\alpha\approx0$, so in this case the condition is simply
$q\le5/6$, which might even be achievable in a binary neutron star system.
Had we used the more exact formula of Eggleton, Eq. (\ref{eggleton}), we would
have found $q\le0.78$.  Note that it has often been assumed that
$R\propto m_{ns}^{-1/3}$ in such discussions~\cite{PZ}, which is equivalent to
$\alpha=-1/3$.  This is unjustified, and results in the upper limit
$q=2/3$ which might inappropriately rule out stable mass transfer in the case
of two neutron stars.

A number of other conditions must hold for stable mass transfer to occur.
First, the orbital separation $a$ at the onset
must exceed the last stable orbit around the black hole, so that $a>6GM_{BH}/c^2$,
or
\begin{equation}
q\ge6{R_\ell\over a}{GM_{BH}\over Rc^2}\,.
\label{agt6m}\end{equation}
Second, the tidal bulge raised on the neutron star must stay outside of the
black hole's Schwarzschild radius.  Kochanek~\cite{Kochanek} gives an estimate
of the height of the tidal bulge needed to achieve the required mass loss rate:
\begin{equation}
{\Delta r\over R}=\Biggl[{-\dot q\over\beta_t(1+q)\Omega}\Biggr]^{1/3}\,,\label{tidal}
\end{equation}
where $\beta_t$ is a dimensionless parameter of order 1 and
$\Omega=G^{1/2}M^{1/2}/a^{3/2}$ is the orbital frequency.  For $\dot q$ we
use the equality in Eq.~(\ref{dotq}), which is equivalent to
\begin{equation}
R_{sh}=2GM_{BH}/c^2\le a-R-\Delta r\,.\label{r-dr}\end{equation} Finally, so
that the assumption of a Roche geometry is valid, it should be possible for
tidal synchronization of the neutron star to be maintained.  Bildstein \&
Cutler~\cite{BC} considered this, and derived an upper limit for the separation
$a_{syn}$ at which tidal synchronization could occur by integrating the maximum
torque on the neutron star as it spirals in from infinity and finding where the
neutron star spin frequency could first equal the orbital frequency.  They find
\begin{equation}
a_{syn}\le{M_{BH}^2m_{ns}^2\over400M^3}\Bigl({R\over m_{ns}}\Bigr)^6\,,\label{synch}\end{equation}
which translates to
\begin{equation}
400\Bigl({GM_{BH}\over Rc^2}\Bigr)^5{a\over R_\ell}{(1+q)^3\over q}\le1\,.\label{tidsyn}
\end{equation}

Next we consider the effect of putting some of the angular momentum into an
accretion disc.  Following the discussion of Ref.~\cite{BC}, we
assume an accretion disc contains an amount of angular momentum that grows at
the rate
\begin{equation}
\dot J_d=-(1-f)M^{3/2}a^{1/2}(1+q)^{-4}\dot q\,,\label{dotjd}\end{equation}
where $f$ is a parameter, taken to be a fit to the numerical results of Hut \&
Paczy\'nski~\cite{HP}:
\begin{equation}
f=5q^{1/3}/3-3q^{2/3}/2\,.\label{hut}\end{equation}
We then find the new condition for angular momentum conservation to be
\begin{equation}
\dot J+\dot J_d=\dot J_{GW}\,,\label{jdot}\end{equation}
which yields
\begin{equation}
\dot q\Biggl[{\alpha\over2}-{1\over6}+{f-q^2\over1+q}\Biggr]\ge
-{32\over5}{G^3\over c^5}{q^2 M^3\over(1+q)a^4}\,.\label{dotqd}\end{equation}
Therefore, the new condition for stable mass transfer is
\begin{equation}
(q^2-f)/(1+q)\le \alpha/2-1/6\,.\label{smtf}\end{equation}
The case $f=1$ corresponds to neglecting the existance of an accretion disc.

It remains to determine when an accretion disc is likely to form.  Initially,
matter flowing from the neutron star to the black hole through the inner
Lagrangian point passes close to the black hole and falls in.  However, as the
neutron star spirals away, the accretion stream trajectory moves outside the
Schwarzschild radius.  When the trajectory doesn't even penetrate the
marginally stable orbit, an accretion disc will begin to form.  Particle
trajectory computations of the Roche geometry by Shore, Livio \& van den
Huevel~\cite{SLv} suggest that its closest approach to the black hole is
\begin{equation}
R_c=a(1+q)(0.5-0.227\ln q)^4\,.
\label{shore}
\end{equation}
Equating $R_c$ to $6GM_{BH}/c^2$ yields
\begin{equation}
(0.5-0.227\ln q)^4(1+q)\ge6{GM_{BH}\over Rc^2}{R_\ell\over a}\,.
\label{disc}
\end{equation}

\begin{figure}[h]
\vspace{24pc}
\includegraphics{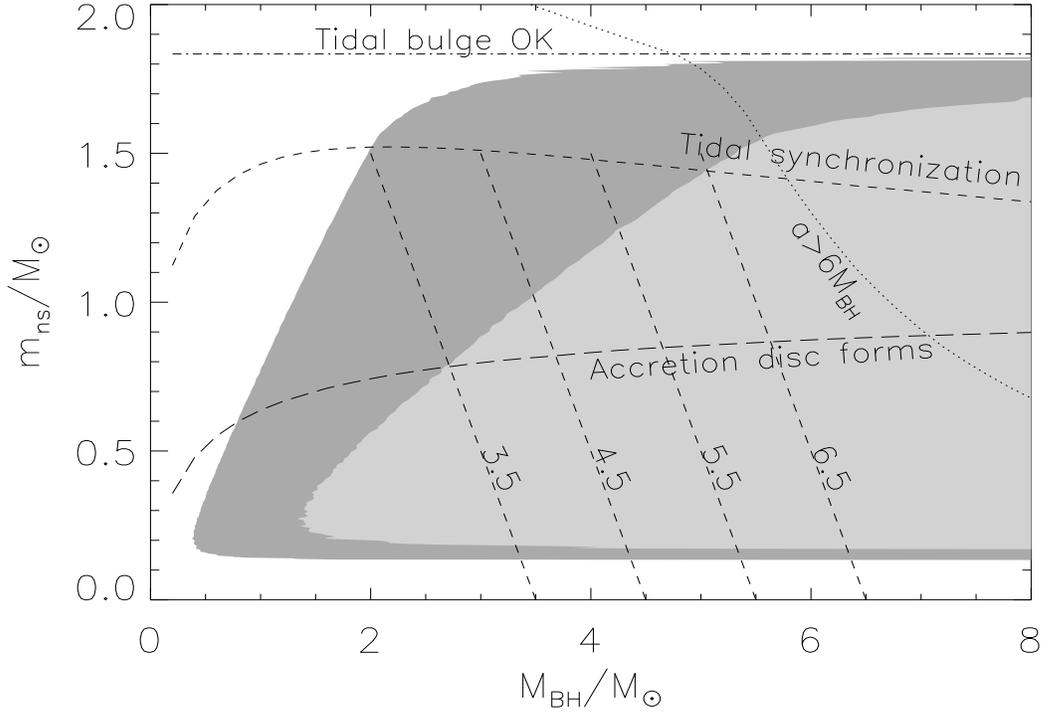}
\caption{The dark and light shaded regions show the binary masses for which
mass transfer in a black hole--neutron star binary will be stable in the
absence of, and the presence of, an accretion disc. The constraints
Eq.~(\ref{agt6m}) ($a>6M_{BH}$), Eq.~(\ref{r-dr}) (Tidal bulge OK),
Eq.~(\ref{tidsyn}) (Tidal synchronization), and Eq.~(\ref{disc})
(Accretion disc
forms) are shown by the appropriately labelled curves. The parallel, diagonal,
dashed lines show evolutionary tracks for the labelled total BH+NS masses,
beginning in each case with $m_{ns}=1.5$ M$_\odot$.}
\label{smt}
\end{figure}

These constraints and allowed regions for stable mass transfer are shown in
Fig.~\ref{smt}.  Apparently, stable mass transfer ceases when
$m_{ns}\approx0.14$ M$_\odot$ if the formation of an accretion disc is ignored.
If the effects of disc formation are included, the stable mass transfer ceases
when $m_{ns}\approx0.22$ M$_\odot$.  In both cases, the neutron star mass
remains above its minimum mass (about 0.09 M$_\odot$ for the equation of state
used here).  Thus, the neutron star does not ``explode'' by reaching its
minimum mass.

\begin{figure}[orbit]
\vspace{22pc}
\includegraphics{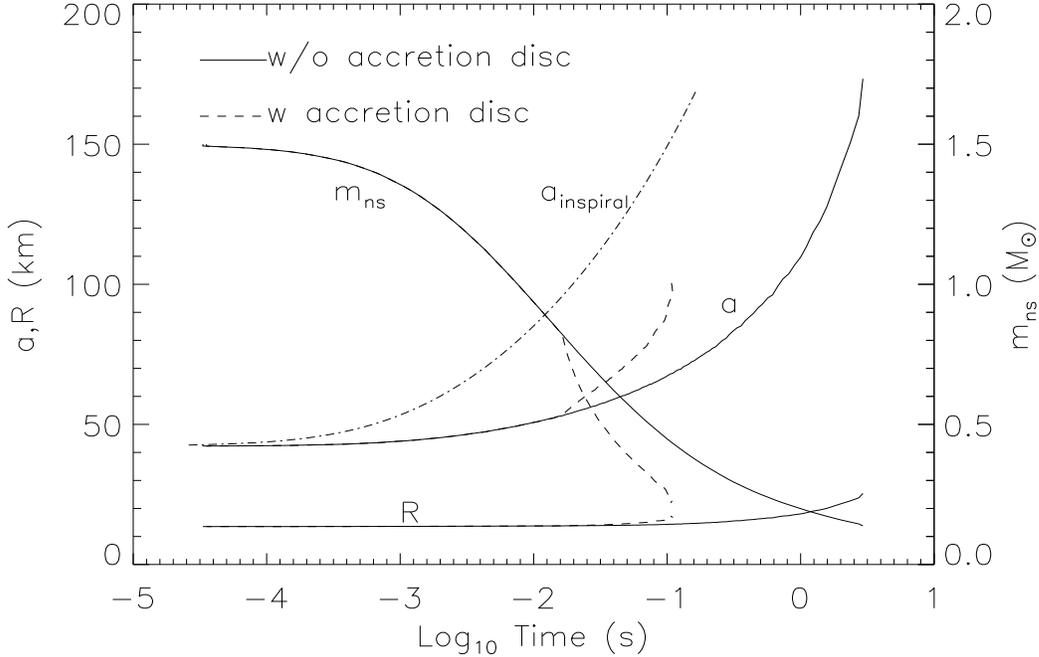}
\caption{The separation of a 1.5 M$_\odot$ neutron star with a 3 M$_\odot$
black hole during a merger is indicated by the dot-dashed line during inspiral
and by a solid line in the outspiral during stable mass transfer.  Other curves
show the neutron star mass and radius during the stable mass transfer
(outspiral) phase.  Solid (dashed) lines are computed by ignoring (including)
the effects of an accretion disc.}
\label{orbit}
\end{figure}

Fig.~\ref{orbit} shows the time development of the orbital separation
$a$ and the neutron star's mass and radius during the inspiral and
stable mass transfer phases.  Solid lines are calculated assuming
there is no accretion disc formed, while dashed lines show the effects
of accretion disc formation.  The time evolutions during stable mass
transfer are obtained from Eq.~(\ref{dotq}) and Eq.~(\ref{dotqd}),
using $\dot m_{ns}=\dot q M/(1+q)^2$.  With disc
formation, the mass transfer is accelerated and the duration of the
stable mass transfer phase is shortened considerably.  Also, the
neutron star spirals out to a smaller radius, and does not
lose as much mass, as in the case when the accretion disc is
ignored.

Therefore, if stable mass transfer can take place, the
timescale over which mass transfer occurs will be much longer than an orbital
period, and lasts perhaps a few tenths of a second.  This is not long enough to
explain gamma-ray bursts.  However, we have also seen the likelihood that an
accretion disc forms is quite large.  Furthermore, the accretion disc extends
to about 100 km.  Even though this is considerably less than Ref.~\cite{PZ}
estimated, the lifetime of such an extended disc is considerable.  To order of
magnitude, it is given by the viscous dissipation time, or
\begin{equation}
\tau_{visc}\sim{D^2\over\alpha c_s H}.
\label{visc}
\end{equation}
Here $D$ is the radial size of the disc, $\alpha$ is the disc's
viscosity parameter,
$c_s$ is the sound speed and $H$ is the disc's thickness.  Note that
$c_s\approx\Omega H$ where $\Omega=2\pi/P=\sqrt{GM_{BH}/D^3}$ is the Kepler
frequency.  Thus,
\begin{equation}
\tau_{visc}\sim {P\over2\pi\alpha} \Biggl({D\over H}\Biggr)^2\,.
\end{equation}
Since the magnitude of $\alpha$ is still undetermined,
and usually quoted~\cite{Bran} to be about 0.01,
and $H$ is likely to be of order $R$, we find $\tau_{visc}\sim 230$ s for our
case.  This alleviates the timescale problem for these models.  Numerical
simulations of such events are in progress, and it remains to be seen if a
viable gamma-ray burst
model from neutron star--black hole coalescence is possible.
If it is, a great deal of the credit should rest with Dave.

We thank Ralph Wijers for discussions
concerning accretion disks.

\end{document}